\documentclass[5p,twocolumn,preprint,12pt]{elsarticle}
\usepackage{graphicx}
\usepackage{hyperref}
\usepackage{color}
\usepackage[table]{xcolor}
\usepackage{boldline,multirow}
\usepackage{booktabs}
\usepackage{pifont}
\usepackage{mathrsfs}
\usepackage[utf8]{inputenc}

\newcommand\Tstrut{\rule{0pt}{2.6ex}}         
\newcommand\Bstrut{\rule[-0.9ex]{0pt}{0pt}}   

\usepackage{color}
\definecolor{rosso}{cmyk}{0,1,1,0.4}
\definecolor{rossos}{cmyk}{0,1,1,0.55}
\definecolor{rossoc}{cmyk}{0,1,1,0.2}
\definecolor{blu}{cmyk}{1,1,0,0.3}
\definecolor{blus}{cmyk}{1,1,0,0.6}
\definecolor{bluc}{cmyk}{1,1,0,0.1}
\definecolor{verde}{cmyk}{0.92,0,0.59,0.25}
\definecolor{verdec}{cmyk}{0.92,0,0.59,0.15}
\definecolor{verdes}{cmyk}{0.92,0,0.59,0.4}
\definecolor{bviolet}{rgb}{0.54, 0.17, 0.89}
\hypersetup{colorlinks,bookmarksopen,bookmarksnumbered,citecolor=bviolet,
linkcolor=bviolet,pdfstartview=FitH,urlcolor=bviolet}

\usepackage{latexsym}
\usepackage{amssymb}
\usepackage{amsmath}
 
 \journal{Nuclear Physics B}
 
\begin{document}

\title{Model-independent Bounds on the Standard Model Effective Theory \\ from Flavour Physics}

\author{Luca Silvestrini}
\ead{luca.silvestrini@roma1.infn.it}
 \address{INFN, Sezione di Roma, P.le A. Moro 2, I-00185 Roma, Italy}
 \address{Theoretical Physics Department, CERN, Geneva, Switzerland}
\author{Mauro Valli}
\ead{mvalli@uci.edu}
 \address{Department of Physics and Astronomy, University of California, Irvine, CA 92697 USA}

\begin{abstract}
  Meson-antimeson mixing provides the most stringent constraints on
  baryon- and lepton-number conserving New Physics, probing scales
  higher than $10^5$ TeV. In the context of the effective theory of
  weak interactions, these constraints translate into severe bounds on
  the coefficients of $\Delta F=2$ operators. Generalizing to the
  effective theory invariant under the Standard Model gauge group,
  valid above the electroweak scale, the bounds from $\Delta F=2$
  processes also affect $\Delta F=1$ and even $\Delta F=0$ operators,
  due to log-enhanced radiative corrections induced by Yukawa
  couplings. We systematically analyze the effect of the
  renormalization group evolution above the electroweak scale and
  provide for the first time the full set of constraints on all
  relevant dimension-six operators.
\end{abstract}


\maketitle

The absence of tree-level flavour changing neutral currents (FCNC) in
the Standard Model (SM), and their suppression at the loop level
\cite{Cabibbo:1963yz,Glashow:1970gm,Kobayashi:1973fv}, make FCNC
processes a very sensitive probe of New Physics (NP). In particular,
meson-antimeson mixing, a FCNC process with flavour quantum number $F$
changed by two units ($\Delta F=2$), provides to date the most
stringent constraints on baryon- and lepton-number conserving NP,
reaching an astonishing NP scale of $\mathcal{O} (10^5)$ TeV for
strongly-interacting NP with arbitrary flavour
structure~\cite{Gabbiani:1996hi,Bona:2007vi,Isidori:2010kg}. This
extraordinary NP sensitivity is due both to the
Glashow-Iliopoulos-Maiani (GIM) mechanism and to the hierarchical
structure of quark masses and mixing angles. Indeed, the bound on the
NP scale can be lowered to a few TeV by requiring Minimal Flavour
Violation (MFV), \textit{i.e.} the absence of new sources of flavour
violation beyond Yukawa couplings
\cite{Buras:2000dm,DAmbrosio:2002vsn}. In the MFV case, the
sensitivity becomes comparable to other indirect probes of NP such as
electroweak (EW) precision observables or Higgs signal strengths, see,
\textit{e.g.}, the recent works in
\cite{deBlas:2016ojx,Ellis:2018gqa}.

If NP arises at scales much higher than the EW one, its leading
effects in the EW and flavour sectors can be parameterized in terms of
dimension-six local operators built of SM fields and invariant under
the SM gauge group. Those operators, together with the SM, form the
so-called Standard Model effective field theory (SMEFT)
\cite{Buchmuller:1985jz,Grzadkowski:2010es}.  Quantum corrections due
to SM interactions induce a renormalization group (RG) running of
SMEFT operators, which can generate $\Delta F=2$ operators
starting from $\Delta F=1$ ones. One may then wonder if mixing onto
$\Delta F=2$ operators implies any relevant bound on $\Delta F = 1 $
ones, to be eventually compared with present constraints from
$\Delta F= 1$ transitions and/or from other EW processes. RG effects
also modify the Yukawa couplings, leading to a mismatch between the
flavour properties of the SMEFT at the NP and EW scales. This effect
is relevant for non-universal operators in the SMEFT, since gauge
invariance prevents a full alignment in flavour space of non-universal
operators involving left-handed doublets, leading unavoidably to FCNC
contributions that depend on the flavour structure at the NP scale. In
this \textit{Letter} we move our first steps towards a deep
investigation of the flavour structure of the SMEFT, focusing on
$\Delta F=2$ transitions, the most sensitive probes of NP in the
flavour sector, and present bounds on all relevant
operators, including all leading RG effects.

\begin{table*}[htb!]
  \centering
  \begin{tabular}{|c|c|c|c|c|c|c|}
    \hline
    & \boldmath$X^{(\textrm{\bf R})}_1$& \boldmath$X^{(\textrm{\bf I})}_1$& \boldmath$X^{(\textrm{\bf R})}_4$& \boldmath$X^{(\textrm{\bf I})}_4$& \boldmath$X^{(\textrm{\bf R})}_5$& \boldmath$X^{(\textrm{\bf I})}_5$\\  [1.01ex]
    \hline
    \boldmath$s \leftrightarrow d$ & $5.9 \cdot 10^{-13}$ & $20.1 \cdot 10^{-16}$ & $4.0 \cdot 10^{-15}$& $1.8 \cdot 10^{-17}$ & $7.5 \cdot 10^{-14}$& $2.0 \cdot 10^{-16}$\\
    \boldmath$c \leftrightarrow u$ & $2.5 \cdot 10^{-13}$ & $7.5 \cdot 10^{-15}$ & $4.2 \cdot 10^{-14}$& $1.3 \cdot 10^{-15}$ & $2.5 \cdot 10^{-13}$& $7.7 \cdot 10^{-15}$\\
    \boldmath$b \leftrightarrow d$ & $9.5 \cdot 10^{-13}$ & $8.0 \cdot 10^{-13}$ & $2.0 \cdot 10^{-13}$& $1.5 \cdot 10^{-13}$ & $5.0 \cdot 10^{-13}$& $4.8 \cdot 10^{-13}$\\
    \boldmath$b \leftrightarrow s$ & $2.0 \cdot 10^{-11}$ & $6.7 \cdot 10^{-12}$ & $5.8 \cdot 10^{-12}$& $2.0 \cdot 10^{-12}$ & $13.2 \cdot 10^{-12}$& $4.5 \cdot 10^{-12}$\\\hline
  \end{tabular}
  \caption{Bounds on $X^{(\textrm{R,I})}_{1,4,5}$ in GeV$^{-2}$ from
    the analysis of refs.~\cite{Bona:2007vi,UTfit2018}. See the text
    for details.}
     \label{tab:DF2}
\end{table*}

\begin{table}[t!]
\begin{tabular}{|cV{3.0}c|c|c|}
  \hline
  &\multicolumn{2}{c|}{{\boldmath${C^{HQ^{(1,3)}}_{ij}}$}
                             [TeV$^{-2}$] } \Tstrut\Bstrut  \\
  [1.01ex]
   {\boldmath$ i j$} \Tstrut\Bstrut& $Y_{D}$ diag & $Y_{U}$ diag \\ [1.01ex]
  \hline
  \textbf{11} \Tstrut\Bstrut & 
 $\varnothing$ &   4.1$^{\,\Box}$ 10$^{-3}$  \\
  \textbf{12} \Tstrut\Bstrut & 
  (8.9$^{\,\Box}$, 3.8$^{\,\Box}$) 10$^{-4}$ &  (9.9$^{\,\Box}$, 3.8$^{\,\Box}$) 10$^{-4}$ \\
  \textbf{13} \Tstrut\Bstrut & 
  (7.4$^{\,\triangle}$, 6.3$^{\,\triangle}$) 10$^{-3}$ & (7.6$^{\,\triangle}$, 6.4$^{\,\triangle}$) 10$^{-3}$ \\
  \textbf{22} \Tstrut\Bstrut & 
  $\varnothing$ &  4.1$^{\,\Box}$ 10$^{-3}$ \\
  \textbf{23} \Tstrut\Bstrut & 
 (3.0$^{\,\bigtriangledown}$, 1.0$^{\,\bigtriangledown}$) 10$^{-2}$ &  (3.1$^{\,\bigtriangledown}$, 1.0$^{\,\bigtriangledown}$) 10$^{-2}$ \\
  \textbf{33} \Tstrut\Bstrut & 
  $ \varnothing $ &  7.3$^{\,\triangle}$ 10$^{-1}$  \\[1.1ex]
  \hline
\end{tabular}
\caption{Constraints on (real, imaginary) parts of Wilson coefficients
  $C^{HQ^{(1,3)}}_{ij}$ obtained from 
  $K-\bar{K}$ ($\Box$), $D-\bar{D}$ ($\Diamond$),
  $B_{d}-\bar{B}_{d}$ ($\triangle$) or
  $B_{s}-\bar{B}_{s}$ ($\bigtriangledown$)
  mixing. Middle and right columns correspond to flavour alignment along the
  down-quark (up-quark) sector. Entries with no bound are denoted
  by $\varnothing$. A single bound is quoted for entries required to
  be real by Hermiticity.}
 \label{tab:HQ}
 \vspace{-1mm}
\end{table}
\begin{table}[t!]
\begin{tabular}{|cV{3.0}c|c|c|}
\hline
\multirow{3}{*}{}& \multicolumn{2}{c|}
{{\boldmath$C^{LeQu}_{ijkl}$} [TeV$^{-2}$] \ \ \ \ {\boldmath$C^{LedQ}_{ijkl}$} [TeV$^{-2}$]} 
\Tstrut\Bstrut  \\ [1.01ex]
 \clineB{2-2}{-0.1}
 {\boldmath$ i j k l $} & $Y_{D}$ diag & $Y_{U}$ diag  \\ [1.01ex]
\hline
\textbf{2221} \Tstrut\Bstrut & 
(5.1$^{\,\Diamond}$, 1.6$^{\,\Diamond}$) 10$^{-1}$ &  
(4.2$^{\,\Box}$, 0.13$^{\,\Box}$) 10$^{-1}$  \\
\textbf{2222} \Tstrut\Bstrut & 
(22$^{\,\Diamond}$, 6.8$^{\,\Diamond}$) 10$^{-1}$ & 
(18$^{\,\Box}$, 0.58$^{\,\Box}$) 10$^{-1}$ \\
\textbf{2223} \Tstrut\Bstrut & 
($\varnothing$, $\varnothing$) & 
(4.3$^{\,\Box}$, 1.6$^{\,\Box}$) \\
\textbf{3321} \Tstrut\Bstrut & 
(3.0$^{\,\Diamond}$, 0.93$^{\,\Diamond}$) 10$^{-2}$ &  
(24$^{\,\Box}$, 0.8$^{\,\Box}$) 10$^{-3}$ \\
 \textbf{3322} \Tstrut\Bstrut & 
(1.3$^{\,\Diamond}$, 0.4$^{\,\Diamond}$) 10$^{-1}$ &  
(10$^{\,\Box}$, 0.34$^{\,\Box}$) 10$^{-2}$ \\
 \textbf{3323} \Tstrut\Bstrut & 
(3.1$^{\,\Diamond}$, 3.6$^{\,\Diamond}$) &  
(2.5$^{\,\Box}$, 0.9$^{\,\Box}$) 10$^{-1}$ \\
 \textbf{3331} \Tstrut\Bstrut & 
($\varnothing$, 9.5$^{\,\Diamond}$) &  
(8.5$^{\,\triangle}$, 11$^{\,\triangle}$)  \\
 \textbf{3332} \Tstrut\Bstrut & 
($\varnothing$, $\varnothing$) &  
($\varnothing$, 8.9$^{\,\bigtriangledown}$)  \\[1.1ex]
\hline
\end{tabular}
\caption{Same as table \ref{tab:HQ} for $C^{LeQu}_{ijkl}$ and
  $C^{LedQ}_{ijkl}$.}
 \label{tab:LE}
 \vspace{-3mm}
\end{table}
Let us begin highlighting the importance of $\Delta F=2$ processes. Extending the notation of~\cite{DAmbrosio:2002vsn}, we define the fundamental FCNC MFV
coupling between generations $i$ and $j$ in the basis of diagonal down
or up Yukawa couplings as
$(\,\lambda_\mathrm{FC}^{(d)}\,)_{i\neq j} \equiv (\,Y_U^{}
Y_U^\dagger \,)_{ij} \sim Y_t^2 V_{3i}^* V_{3j}$ or
$(\,\lambda_\mathrm{FC}^{(u)}\,)_{i\neq j} \equiv (Y_D^{} Y_D^\dagger
\,)_{ij} \sim Y_b^2 V_{3i} V_{3j}^*$ respectively; $i,j=1,2,3$ are
flavour indices, $Y_{q}$ the Yukawa coupling for quark $q$ in the
diagonal basis and $V$ the CKM matrix. We characterize the typical SM
FCNC scale as
$\Lambda_{0} \equiv Y_t \sin^2 \theta_W M_W/\alpha\sim 2.3~$TeV.  An
MFV-type NP model will generate $\Delta F=1$ and $\Delta F=2$
operators with the same chiral structure as the SM contribution, up to
corrections proportional to further powers of Yukawa couplings, with
coefficients of
$\mathcal{O}((\lambda_\mathrm{FC}^{(d),(u)})_{ij}^2/\Lambda^2)$ for
$\Delta F=2$ and of
$\mathcal{O}((\lambda_\mathrm{FC}^{(d),(u)})_{ij}/\Lambda^2)$ for
$\Delta F=1$ processes in the down and up sector respectively. In the
down sector, those have the same structure of top-mediated SM
contributions, so the most stringent constraints are expected from
top-dominated processes such as meson-antimeson mixing or
$b \to s \gamma$, leading to lower bounds on the NP scale of
few TeV~\cite{Buras:2000dm,DAmbrosio:2002vsn}.
\begin{table}[b!]
\resizebox{\columnwidth}{!}{
\begin{tabular}{|cV{3.0}c|c|c|c|c|}
\hline
\multirow{3}{*}{}& \multicolumn{2}{c|}{{\boldmath$C^{ud^{(1)}}_{ijkl}$} [TeV$^{-2}$]} \Tstrut\Bstrut  & 
\multicolumn{2}{c|}{{\boldmath$C^{ud^{(8)}}_{ijkl}$} [TeV$^{-2}$]}  \\ [1.01ex]
 \clineB{2-5}{-0.1}
 {\boldmath$ i j k l $} & $Y_{D}$ diag & $Y_{U}$ diag & $Y_{D}$ diag & $Y_{U}$ diag \\ [1.01ex]
\hline
\textbf{1112} \Tstrut\Bstrut & 
($\varnothing$, 1.1$^{\,\Box}$) &  
($\varnothing$, $\varnothing$) & 
($\varnothing$, 0.10$^{\,\Box}$)  &
($\varnothing$, $\varnothing$) \\
\textbf{1212} \Tstrut\Bstrut & 
($\varnothing$, 2.5$^{\,\Box}$) 10$^{-1}$  &  
($\varnothing$, 2.5$^{\,\Box}$) 10$^{-1}$  & 
(99$^{\,\Box}$, 0.45$^{\,\Box}$) 10$^{-1}$  &  
(99$^{\,\Box}$, 0.45$^{\,\Box}$) 10$^{-1}$\\
\textbf{1213} \Tstrut\Bstrut & 
($\varnothing$, $\varnothing$)   &  
($\varnothing$, $\varnothing$)   & 
($\varnothing$, 7.0$^{\,\Diamond}$)   &  
($\varnothing$, $\varnothing$) \\
\textbf{1221} \Tstrut\Bstrut & 
(360$^{\,\Box}$, 0.95$^{\,\Box}$) 10$^{-2}$  & 
($\varnothing$, 4.6$^{\,\Box}$) & 
(38$^{\,\Box}$, 0.17$^{\,\Box}$) 10$^{-2}$  & 
($\varnothing$, 8.3$^{\,\Box}$) 10$^{-1}$\\
\textbf{1222} \Tstrut\Bstrut & 
($\varnothing$, 11$^{\,\Diamond}$)   & 
($\varnothing$, 11$^{\,\Diamond}$) & 
($\varnothing$, 3.6$^{\,\Diamond}$)  & 
($\varnothing$, 3.6$^{\,\Diamond}$) \\
\textbf{1223} \Tstrut\Bstrut & 
($\varnothing$, 4.7$^{\,\Diamond}$)   & 
($\varnothing$, $\varnothing$) & 
($\varnothing$, 1.6$^{\,\Diamond}$)  & 
($\varnothing$, $\varnothing$) \\
\textbf{1231} \Tstrut\Bstrut & 
(2.4$^{\,\triangle}$, 2.3$^{\,\triangle}$) &  
($\varnothing$, $\varnothing$) & 
(1.9$^{\,\triangle}$, 1.4$^{\,\triangle}$) & 
($\varnothing$, $\varnothing$) \\
\textbf{1232} \Tstrut\Bstrut & 
(12$^{\,\bigtriangledown}$, 5.0$^{\,\bigtriangledown}$) &  
($\varnothing$, $\varnothing$)  & 
(4.6$^{\,\Diamond}$, 4.5$^{\,\bigtriangledown}$)  & 
(4.6$^{\,\Diamond}$, 10$^{\,\Diamond}$)\\
\textbf{1233} \Tstrut\Bstrut & 
(6.0$^{\,\Diamond}$, $\varnothing$) &  
($\varnothing$, $\varnothing$)  & 
(2.0$^{\,\Diamond}$, 4.5$^{\,\Diamond}$)  & 
($\varnothing$, $\varnothing$) \\
\textbf{1312} \Tstrut\Bstrut & 
($\varnothing$, 5.7$^{\,\Box}$)  &  
(11$^{\,\Box}$, 0.21$^{\,\Box}$) 10$^{-1}$ & 
($\varnothing$, 1.0$^{\,\Box}$)  & 
(21$^{\,\Box}$, 0.37$^{\,\Box}$) 10$^{-2}$\\
\textbf{1313} \Tstrut\Bstrut & 
(2.2$^{\,\triangle}$, 2.1$^{\,\triangle}$)  &  
(2.2$^{\,\triangle}$, 2.1$^{\,\triangle}$) & 
(1.7$^{\,\triangle}$, 1.3$^{\,\triangle}$) & 
(1.7$^{\,\triangle}$, 1.3$^{\,\triangle}$)\\
\textbf{1321} \Tstrut\Bstrut & 
(2.3$^{\,\Box}$, 0.96$^{\,\Box}$) 10$^{-3}$ & 
(12$^{\,\Box}$, 4.7 $^{\,\Box}$) 10$^{-1}$ & 
(4.2$^{\,\Box}$, 1.7 $^{\,\Box}$) 10$^{-4}$ &
(2.0$^{\,\Box}$, 0.84 $^{\,\Box}$) 10$^{-1}$ \\
\textbf{1331} \Tstrut\Bstrut & 
(2.1$^{\,\triangle}$, 2.0$^{\,\triangle}$) 10$^{-1}$ & 
($\varnothing$, $\varnothing$) & 
(1.7$^{\,\triangle}$, 1.2$^{\,\triangle}$) 10$^{-1}$ &
($\varnothing$, $\varnothing$) \\
\textbf{1332} \Tstrut\Bstrut & 
(1.0$^{\,\bigtriangledown}$, 4.3 $^{\,\bigtriangledown}$) 10$^{-1}$ & 
($\varnothing$, $\varnothing$) & 
(8.9$^{\,\bigtriangledown}$, 3.8 $^{\,\bigtriangledown}$) 10$^{-1}$ &
($\varnothing$, $\varnothing$) \\
\textbf{2212} \Tstrut\Bstrut & 
(83$^{\,\Box}$, 0.22$^{\,\Box}$) 10$^{-2}$ & 
(83$^{\,\Box}$, 0.22$^{\,\Box}$) 10$^{-2}$ & 
(89$^{\,\Box}$, 0.4$^{\,\Box}$) 10$^{-3}$ &
(89$^{\,\Box}$, 0.4$^{\,\Box}$) 10$^{-3}$ \\
\textbf{2213} \Tstrut\Bstrut & 
(4.7$^{\,\triangle}$, 4.5 $^{\,\triangle}$) 10$^{-1}$ & 
($\varnothing$, $\varnothing$)  & 
(3.7$^{\,\triangle}$, 2.8 $^{\,\triangle}$) 10$^{-1}$ &
($\varnothing$, 11$^{\,\Box}$) \\
\textbf{2223} \Tstrut\Bstrut & 
( 28$^{\,\bigtriangledown}$,  9.7$^{\,\bigtriangledown}$) 10$^{-1}$ & 
($\varnothing$, $\varnothing$)  & 
( 25$^{\,\bigtriangledown}$,  8.6$^{\,\bigtriangledown}$) 10$^{-1}$ & 
($\varnothing$, $\varnothing$) \\
\textbf{2312} \Tstrut\Bstrut & 
($\varnothing$, 5.1$^{\,\Box}$) 10$^{-2}$ & 
( 11$^{\,\Box}$, 0.19$^{\,\Box}$) 10$^{-3}$ & 
( 200$^{\,\Box}$, 0.92$^{\,\Box}$) 10$^{-2}$ &
( 11$^{\,\Box}$, 0.33$^{\,\Box}$) 10$^{-4}$  \\
\textbf{2313} \Tstrut\Bstrut & 
(1.9$^{\,\triangle}$, 1.9$^{\,\triangle}$) 10$^{-2}$ &
(1.9$^{\,\triangle}$, 1.9$^{\,\triangle}$) 10$^{-2}$ &
(1.5$^{\,\triangle}$, 1.2$^{\,\triangle}$) 10$^{-2}$ &
(1.5$^{\,\triangle}$, 1.2$^{\,\triangle}$) 10$^{-2}$ \\
\textbf{2321} \Tstrut\Bstrut & 
(5.5$^{\,\Box}$, 2.2$^{\,\Box}$) 10$^{-4}$ & 
(5.5$^{\,\Box}$, 2.2$^{\,\Box}$) 10$^{-4}$ & 
(9.9$^{\,\Box}$, 4.0$^{\,\Box}$) 10$^{-5}$ &
(9.9$^{\,\Box}$, 4.0$^{\,\Box}$) 10$^{-5}$ \\
\textbf{2323} \Tstrut\Bstrut & 
(1.2$^{\,\bigtriangledown}$, 0.40$^{\,\bigtriangledown}$) 10$^{-1}$ & 
(1.2$^{\,\bigtriangledown}$, 0.40$^{\,\bigtriangledown}$) 10$^{-1}$ & 
(1.0$^{\,\bigtriangledown}$, 0.36$^{\,\bigtriangledown}$) 10$^{-1}$ &
(1.0$^{\,\bigtriangledown}$, 0.36$^{\,\bigtriangledown}$) 10$^{-1}$ \\
\textbf{2331} \Tstrut\Bstrut & 
(4.7$^{\,\triangle}$, 4.5$^{\,\triangle}$) 10$^{-2}$ & 
($\varnothing$, 6.0 $^{\,\Box}$) & 
(3.8$^{\,\triangle}$, 2.8 $^{\,\triangle}$) 10$^{-2}$ &
(2.5$^{\,\Box}$, 1.1$^{\,\Box}$)  \\
\textbf{2332} \Tstrut\Bstrut & 
(2.4$^{\,\bigtriangledown}$, 0.82$^{\,\bigtriangledown}$) 10$^{-1}$ & 
($\varnothing$, $\varnothing$)  & 
(2.1$^{\,\bigtriangledown}$, 0.72$^{\,\bigtriangledown}$) 10$^{-1}$ & 
($\varnothing$, $\varnothing$) \\
\textbf{3311} \Tstrut\Bstrut & 
($\varnothing$, $\varnothing$) & 
($\varnothing$, $\varnothing$) & 
($\varnothing$, $\varnothing$) & 
(4.8$^{\,\Box}$, $\varnothing$) \\
\textbf{3312} \Tstrut\Bstrut & 
(13$^{\,\Box}$, 5.1$^{\,\Box}$) 10$^{-3}$ & 
(4.4$^{\,\Box}$, 2.0$^{\,\Box}$) 10$^{-5}$ & 
(2.3$^{\,\Box}$, 0.92$^{\,\Box}$) 10$^{-3}$ & 
(8.0$^{\,\Box}$, 3.4$^{\,\Box}$) 10$^{-6}$ \\
\textbf{3313} \Tstrut\Bstrut & 
(2.0$^{\,\triangle}$, 1.9$^{\,\triangle}$) 10$^{-3}$ & 
(2.0$^{\,\triangle}$, 1.9$^{\,\triangle}$) 10$^{-3}$ & 
(1.6$^{\,\triangle}$, 1.2$^{\,\triangle}$) 10$^{-3}$ & 
(1.6$^{\,\triangle}$, 1.2$^{\,\triangle}$) 10$^{-3}$ \\
\textbf{3322} \Tstrut\Bstrut & 
($\varnothing$, $\varnothing$) & 
($\varnothing$, $\varnothing$) & 
($\varnothing$, $\varnothing$) & 
(4.8$^{\,\Box}$, $\varnothing$) \\
\textbf{3323} \Tstrut\Bstrut & 
(10$^{\,\bigtriangledown}$, 3.4 $^{\,\bigtriangledown}$) 10$^{-3}$ & 
(10$^{\,\bigtriangledown}$, 3.4 $^{\,\bigtriangledown}$) 10$^{-3}$ &
(8.7$^{\,\bigtriangledown}$, 3.0 $^{\,\bigtriangledown}$) 10$^{-3}$ & 
(8.7$^{\,\bigtriangledown}$, 3.0 $^{\,\bigtriangledown}$) 10$^{-3}$ \\[1.1ex]
\hline
\end{tabular}}
 \caption{Same as table \ref{tab:HQ} for $C^{ud^{(1)}}_{ijkl}$ and $C^{ud^{(8)}}_{ijkl}$.}
 \label{tab:ud}
 \vspace{-3mm}
\end{table}
\begin{table}[t!]
\resizebox{.916\columnwidth}{!}{
\begin{tabular}{|cV{3.0}c|c|c|c|c|}
\hline
\multirow{3}{*}{}& \multicolumn{2}{c|}{{\boldmath$C^{QuQd^{(1)}}_{ijkl}$} [TeV$^{-2}$]} \Tstrut\Bstrut  & 
\multicolumn{2}{c|}{{\boldmath$C^{QuQd^{(8)}}_{ijkl}$} [TeV$^{-2}$]}  \\ [1.01ex]
 \clineB{2-5}{-0.1}
 {\boldmath$ijkl$} & $Y_{D}$ diag & $Y_{U}$ diag & $Y_{D}$ diag & $Y_{U}$ diag \\ [1.01ex]
\hline
\textbf{1111} \Tstrut\Bstrut & 
 ($\varnothing$, $\varnothing$) &  
($\varnothing$, 4.8$^{\,\Box}$)  & 
($\varnothing$, $\varnothing$) & 
($\varnothing$, $\varnothing$) \\
\textbf{1112} \Tstrut\Bstrut & 
 (92$^{\,\Box}$, 0.41$^{\,\Box}$) 10$^{-1}$ &  
(9.5$^{\,\Box}$, 0.31$^{\,\Box}$)  & 
 ($\varnothing$, 0.49$^{\,\Box}$)  &  
($\varnothing$, 1.6$^{\,\Box}$) \\
\textbf{1113} \Tstrut\Bstrut & 
(4.0$^{\,\Diamond}$, 8.9$^{\,\Diamond}$) &  
($\varnothing$, $\varnothing$) & 
($\varnothing$, $\varnothing$) &
($\varnothing$, $\varnothing$) \\
\textbf{1121} \Tstrut\Bstrut & 
($\varnothing$, 3.2$^{\,\Diamond}$)  &   
($\varnothing$, 1.1$^{\,\Box}$)  & 
($\varnothing$, $\varnothing$)  & 
($\varnothing$, $\varnothing$)   \\
\textbf{1122} \Tstrut\Bstrut & 
($\varnothing$, 0.71$^{\,\Diamond}$)   &   
($\varnothing$, 1.6$^{\,\Box}$)   & 
($\varnothing$, 8.6$^{\,\Diamond}$)   & 
($\varnothing$, $\varnothing$)  \\
\textbf{1123} \Tstrut\Bstrut & 
(9.2$^{\,\Diamond}$, 21$^{\,\Diamond}$) 10$^{-1}$  &   
($\varnothing$, $\varnothing$) & 
(11$^{\,\Diamond}$, $\varnothing$) & 
($\varnothing$, $\varnothing$) \\
\textbf{1211} \Tstrut\Bstrut & 
(5.1$^{\,\Diamond}$, 1.6$^{\,\Diamond}$) &  
(95$^{\,\Box}$, 0.43$^{\,\Box}$) 10$^{-1}$ & 
($\varnothing$, 8.3$^{\,\Diamond}$) &
($\varnothing$, 5.1$^{\,\Box}$) 10$^{-1}$  \\
\textbf{1212} \Tstrut\Bstrut & 
(210$^{\,\Box}$, 0.96$^{\,\Box}$) 10$^{-2}$ &  
(22$^{\,\Box}$,  0.1$^{\,\Box}$) 10$^{-1}$ & 
($\varnothing$,  1.1$^{\,\Box}$) 10$^{-1}$ &  
($\varnothing$,  1.2$^{\,\Box}$) 10$^{-1}$ \\
\textbf{1213} \Tstrut\Bstrut & 
(8.9$^{\,\triangle}$, 3.6$^{\,\Diamond}$) &  
($\varnothing$, $\varnothing$) & 
($\varnothing$, $\varnothing$) & 
($\varnothing$, $\varnothing$) \\
\textbf{1221} \Tstrut\Bstrut & 
($\varnothing$, 11$^{\,\Diamond}$) &   
(220$^{\,\Box}$, 0.99$^{\,\Box}$) 10$^{-2}$ & 
($\varnothing$, $\varnothing$) &   
($\varnothing$, 1.2$^{\,\Box}$) 10$^{-1}$  \\
\textbf{1222} \Tstrut\Bstrut & 
(3.1$^{\,\Diamond}$, 2.3$^{\,\Diamond}$) 10$^{-1}$ &  
(14$^{\,\Box}$, 0.4$^{\,\Box}$) 10$^{-2}$ &
($\varnothing$, 9.8$^{\,\Diamond}$) 10$^{-1}$ &  
(10$^{\,\Box}$, 0.53$^{\,\Box}$) 10$^{-2}$ \\
\textbf{1223} \Tstrut\Bstrut & 
(22$^{\,\Diamond}$, 0.69$^{\,\Diamond}$) 10$^{-1}$ &   
($\varnothing$, $\varnothing$) &  
($\varnothing$, 4.3$^{\,\Diamond}$) 10$^{-1}$ &   
($\varnothing$, $\varnothing$)  \\
\textbf{1231} \Tstrut\Bstrut & 
(3.4$^{\,\Diamond}$, 1.4$^{\,\Diamond}$)  &  
($\varnothing$, 0.23$^{\,\Box}$)  & 
($\varnothing$, 8.2$^{\,\Diamond}$) & 
($\varnothing$, 2.8$^{\,\Box}$) \\
\textbf{1232} \Tstrut\Bstrut & 
(4.0$^{\,\Diamond}$, 1.7$^{\,\Diamond}$) 10$^{-2}$ &  
(6.5$^{\,\Box}$, 15$^{\,\Box}$) 10$^{-2}$  & 
(2.4$^{\,\Diamond}$, 1.1$^{\,\Diamond}$) 10$^{-1}$ & 
(1.2$^{\,\Box}$, 2.6$^{\,\Box}$) 10$^{-2}$ \\
\textbf{1233} \Tstrut\Bstrut & 
(9.0$^{\,\Diamond}$, 1.5$^{\,\Diamond}$) 10$^{-3}$ &  
($\varnothing$, $\varnothing$)  & 
(11$^{\,\Diamond}$, 4.4$^{\,\Diamond}$) 10$^{-2}$ & 
($\varnothing$, $\varnothing$) \\
\textbf{1311} \Tstrut\Bstrut & 
($\varnothing$, $\varnothing$) &  
(11$^{\,\Box}$, 4.3$^{\,\Box}$) 10$^{-3}$  & 
($\varnothing$, $\varnothing$) & 
(13$^{\,\Box}$, 5.2$^{\,\Box}$) 10$^{-2}$ \\
\textbf{1312} \Tstrut\Bstrut & 
($\varnothing$, 2.2$^{\,\Box}$) 10$^{-1}$&  
(470$^{\,\Box}$, 8.5$^{\,\Box}$) 10$^{-4}$  & 
($\varnothing$, 2.6$^{\,\Box}$) &  
(5.6$^{\,\Box}$, 0.1$^{\,\Box}$) 10$^{-1}$ \\
\textbf{1313} \Tstrut\Bstrut & 
(3.7$^{\,\triangle}$, 2.8$^{\,\triangle}$) 10$^{-1}$ &  
(3.9$^{\,\triangle}$, 2.9$^{\,\triangle}$) 10$^{-1}$ &  
(2.2$^{\,\triangle}$, 2.1$^{\,\triangle}$) &  
(2.4$^{\,\triangle}$, 2.3$^{\,\triangle}$) \\
\textbf{1321} \Tstrut\Bstrut & 
($\varnothing$, $\varnothing$) &  
(2.4$^{\,\Box}$, 0.11$^{\,\Box}$) 10$^{-3}$  & 
($\varnothing$, $\varnothing$) & 
(2.9$^{\,\Box}$, 1.2$^{\,\Box}$) 10$^{-2}$ \\
\textbf{1322} \Tstrut\Bstrut & 
($\varnothing$, $\varnothing$) &  
(21$^{\,\Box}$, 0.37$^{\,\Box}$) 10$^{-2}$  & 
($\varnothing$, $\varnothing$) & 
(41$^{\,\Box}$, 0.74$^{\,\Box}$) 10$^{-3}$   \\
\textbf{1323} \Tstrut\Bstrut & 
($\varnothing$, $\varnothing$) &  
(4.5$^{\,\triangle}$, 4.3$^{\,\triangle}$) 10$^{-1}$  & 
($\varnothing$, $\varnothing$) & 
(3.4$^{\,\triangle}$, 2.6$^{\,\triangle}$) 10$^{-1}$ \\
\textbf{1331} \Tstrut\Bstrut & 
($\varnothing$, $\varnothing$) &  
(5.5$^{\,\Box}$, 2.3$^{\,\Box}$) 10$^{-2}$  & 
($\varnothing$, $\varnothing$) & 
(6.7$^{\,\Box}$, 2.8$^{\,\Box}$) 10$^{-1}$ \\
\textbf{1332} \Tstrut\Bstrut & 
($\varnothing$, $\varnothing$) &  
(5.0$^{\,\Box}$, 0.16$^{\,\Box}$) 10$^{-4}$  & 
($\varnothing$, $\varnothing$) & 
(5.9$^{\,\Box}$, 0.12$^{\,\Box}$) 10$^{-4}$  \\
\textbf{1333} \Tstrut\Bstrut & 
($\varnothing$, $\varnothing$) &  
(1.7$^{\,\triangle}$, 2.3$^{\,\triangle}$) 10$^{-1}$  & 
($\varnothing$, $\varnothing$) & 
(2.0$^{\,\triangle}$, 2.7$^{\,\triangle}$) 10$^{-1}$  \\
 \textbf{2111} \Tstrut\Bstrut & 
($\varnothing$, 3.15$^{\,\Diamond}$)   &
($\varnothing$, 1.1$^{\,\Box}$) &  
($\varnothing$, $\varnothing$)   &
($\varnothing$, $\varnothing$)  \\
 \textbf{2112} \Tstrut\Bstrut & 
($\varnothing$, 7.1$^{\,\Diamond}$) 10$^{-1}$  &
($\varnothing$, 3.2$^{\,\Box}$)  & 
($\varnothing$, 8.6$^{\,\Diamond}$)  &
($\varnothing$, 9.4$^{\,\Box}$)  \\
 \textbf{2113} \Tstrut\Bstrut & 
(9.2$^{\,\Diamond}$, 21$^{\,\Diamond}$) 10$^{-1}$  &
($\varnothing$, $\varnothing$) &  
(11$^{\,\Diamond}$, $\varnothing$)  &
($\varnothing$, $\varnothing$) \\
\textbf{2121} \Tstrut\Bstrut & 
($\varnothing$, 2.4$^{\,\Box}$) 10$^{-1}$ &
($\varnothing$, 2.6$^{\,\Box}$) 10$^{-1}$ &  
($\varnothing$, 2.9$^{\,\Box}$)  &
($\varnothing$, 3.1$^{\,\Box}$)  \\
\textbf{2122} \Tstrut\Bstrut & 
(5.3$^{\,\Diamond}$, 0.17$^{\,\Diamond}$) &
(5.1$^{\,\Diamond}$, 0.16$^{\,\Diamond}$) &  
($\varnothing$, 2.0$^{\,\Diamond}$) &
($\varnothing$, 1.9$^{\,\Diamond}$)  \\
\textbf{2123} \Tstrut\Bstrut & 
(2.1$^{\,\Diamond}$, 4.7$^{\,\Diamond}$) 10$^{-1}$ &
(2.0$^{\,\Diamond}$, 4.5$^{\,\Diamond}$) 10$^{-1}$ &
(2.6$^{\,\Diamond}$, 5.7$^{\,\Diamond}$) &
(2.4$^{\,\Diamond}$, 5.4$^{\,\Diamond}$)  \\
 \textbf{2131} \Tstrut\Bstrut & 
($\varnothing$, $\varnothing$) &
($\varnothing$, 6.0$^{\,\Box}$)  &
($\varnothing$, $\varnothing$) & 
($\varnothing$, $\varnothing$)  \\
 \textbf{2132} \Tstrut\Bstrut & 
($\varnothing$, 3.7$^{\,\Diamond}$) &
($\varnothing$, $\varnothing$)  &
($\varnothing$, $\varnothing$) & 
($\varnothing$, $\varnothing$)  \\
 \textbf{2133} \Tstrut\Bstrut & 
(4.8$^{\,\Diamond}$, 10$^{\,\Diamond}$) &
($\varnothing$, $\varnothing$) &
($\varnothing$, $\varnothing$) &
($\varnothing$, $\varnothing$) \\
\textbf{2211} \Tstrut\Bstrut & 
($\varnothing$, 8.0$^{\,\Diamond}$) &
(22$^{\,\Box}$, 0.1$^{\,\Box}$) 10$^{-1}$ &  
($\varnothing$, $\varnothing$) &
($\varnothing$, 1.2$^{\,\Box}$) 10$^{-1}$ \\
\textbf{2212} \Tstrut\Bstrut & 
(1.9$^{\,\Diamond}$, 0.527$^{\,\Diamond}$) &
(2.3$^{\,\Box}$, 0.11$^{\,\Box}$) 10$^{-2}$ &  
(1.4$^{\,\Diamond}$, 1.3$^{\,\Diamond}$) &
(28$^{\,\Box}$, 0.13$^{\,\Box}$) 10$^{-1}$ \\
\textbf{2213} \Tstrut\Bstrut & 
($\varnothing$, 8.6$^{\,\Diamond}$) 10$^{-1}$ & 
(8.0$^{\,\triangle}$, 10$^{\,\triangle}$) &
(9.0$^{\,\Diamond}$, 0.28$^{\,\Diamond}$) &
($\varnothing$, $\varnothing$) \\
\textbf{2221} \Tstrut\Bstrut & 
(48$^{\,\Box}$, 0.22$^{\,\Box}$) 10$^{-2}$ &
(51$^{\,\Box}$, 0.23$^{\,\Box}$) 10$^{-2}$ &
(58$^{\,\Box}$, 0.26$^{\,\Box}$) 10$^{-1}$ &
(61$^{\,\Box}$, 0.27$^{\,\Box}$) 10$^{-1}$ \\
\textbf{2222} \Tstrut\Bstrut & 
(12$^{\,\Diamond}$, 7.0$^{\,\Diamond}$) 10$^{-1}$ &
(8.3$^{\,\Box}$, 0.38$^{\,\Box}$) 10$^{-2}$ & 
(6.1$^{\,\Diamond}$, 2.4$^{\,\Diamond}$)  & 
(4.3$^{\,\Box}$, 0.16$^{\,\Box}$) 10$^{-1}$ \\
\textbf{2223} \Tstrut\Bstrut & 
(9.7$^{\,\Diamond}$, 0.30$^{\,\Diamond}$) &
($\varnothing$, 7.1$^{\,\bigtriangledown}$) & 
($\varnothing$, 3.4$^{\,\Diamond}$)  & 
($\varnothing$, $\varnothing$) \\
\textbf{2231} \Tstrut\Bstrut & 
($\varnothing$, 2.2$^{\,\Box}$) &
(31$^{\,\Box}$, 0.53$^{\,\Box}$) 10$^{-1}$ &
($\varnothing$, $\varnothing$)  &
($\varnothing$, 0.64$^{\,\Box}$) \\
\textbf{2232} \Tstrut\Bstrut & 
(1.8$^{\,\Diamond}$, 0.72$^{\,\Diamond}$) 10$^{-1}$ &  
(2.8$^{\,\Box}$, 1.3$^{\,\Box}$) 10$^{-1}$ & 
(11$^{\,\Diamond}$, 4.3$^{\,\Diamond}$) 10$^{-1}$ &
(5.0$^{\,\Box}$, 11$^{\,\Box}$) 10$^{-2}$  \\
 \textbf{2233} \Tstrut\Bstrut & 
(3.9$^{\,\Diamond}$, 0.64$^{\,\Diamond}$) 10$^{-2}$ &  
($\varnothing$, $\varnothing$)& 
(4.7$^{\,\Diamond}$, 1.9$^{\,\Diamond}$) 10$^{-1}$ &
($\varnothing$, $\varnothing$) \\
\textbf{2311} \Tstrut\Bstrut & 
($\varnothing$, $\varnothing$)& 
(2.4$^{\,\Box}$, 1.1$^{\,\Box}$) 10$^{-3}$ & 
($\varnothing$, $\varnothing$) & 
(2.9$^{\,\Box}$, 1.2$^{\,\Box}$) 10$^{-2}$ \\
\textbf{2312} \Tstrut\Bstrut & 
($\varnothing$, $\varnothing$) &  
(10$^{\,\Box}$, 0.2$^{\,\Box}$) 10$^{-3}$ & 
($\varnothing$, $\varnothing$) &  
(5.9$^{\,\Box}$, 0.1$^{\,\Box}$) 10$^{-2}$ \\
\textbf{2313} \Tstrut\Bstrut & 
($\varnothing$, $\varnothing$) &  
(8.9$^{\,\triangle}$, 6.7$^{\,\triangle}$) 10$^{-2}$ & 
($\varnothing$, $\varnothing$) &  
(4.4$^{\,\triangle}$, 3.7$^{\,\triangle}$) 10$^{-1}$ \\
\textbf{2321} \Tstrut\Bstrut & 
(5.3$^{\,\Box}$, 2.2$^{\,\Box}$) 10$^{-4}$ & 
(5.7$^{\,\Box}$, 2.3$^{\,\Box}$) 10$^{-4}$ & 
(6.4$^{\,\Box}$, 2.6$^{\,\Box}$) 10$^{-3}$ & 
(6.8$^{\,\Box}$, 2.8$^{\,\Box}$) 10$^{-3}$ \\
\textbf{2322} \Tstrut\Bstrut & 
($\varnothing$, $\varnothing$) &  
(48$^{\,\Box}$, 0.83$^{\,\Box}$) 10$^{-3}$ &
($\varnothing$, $\varnothing$) &  
(57$^{\,\Box}$, 1.0$^{\,\Box}$) 10$^{-2}$ \\
\textbf{2323} \Tstrut\Bstrut & 
(5.7$^{\,\bigtriangledown}$, 2.0$^{\,\bigtriangledown}$) 10$^{-1}$ &  
(3.9$^{\,\triangle}$, 2.1$^{\,\bigtriangledown}$) 10$^{-1}$ & 
(3.1$^{\,\bigtriangledown}$, 1.1$^{\,\bigtriangledown}$)& 
(2.3$^{\,\triangle}$, 1.1$^{\,\bigtriangledown}$) \\
\textbf{2331} \Tstrut\Bstrut & 
(5.3$^{\,\Box}$, 2.3$^{\,\Box}$) 10$^{-1}$ &  
(13$^{\,\Box}$, 5.4$^{\,\Box}$) 10$^{-3}$ & 
(6.4$^{\,\Box}$, 2.7$^{\,\Box}$) &
(15$^{\,\Box}$, 6.5$^{\,\Box}$) 10$^{-2}$ \\
\textbf{2332} \Tstrut\Bstrut & 
($\varnothing$, $\varnothing$) &
(25$^{\,\Box}$, 0.7$^{\,\Box}$) 10$^{-4}$ &
($\varnothing$, $\varnothing$) &
(25$^{\,\Box}$, 0.51$^{\,\Box}$) 10$^{-4}$  \\
\textbf{2333} \Tstrut\Bstrut & 
 ($\varnothing$, $\varnothing$) &  
(5.0$^{\,\bigtriangledown}$, 1.7$^{\,\bigtriangledown}$) 10$^{-1}$ & 
 ($\varnothing$, $\varnothing$) & 
(6.2$^{\,\bigtriangledown}$, 2.1$^{\,\bigtriangledown}$) 10$^{-1}$ \\
\textbf{3112} \Tstrut\Bstrut & 
 ($\varnothing$, $\varnothing$) & 
 (6.2$^{\,\Box}$, $\varnothing$) & 
 ($\varnothing$, $\varnothing$) &
 ($\varnothing$, $\varnothing$)  \\
\textbf{3121} \Tstrut\Bstrut & 
  ($\varnothing$, $\varnothing$) & 
 ($\varnothing$, 6.0$^{\,\Box}$) & 
 ($\varnothing$, $\varnothing$) &
 ($\varnothing$, $\varnothing$)  \\
\textbf{3122} \Tstrut\Bstrut & 
  ($\varnothing$, 3.7$^{\,\Diamond}$) & 
 ($\varnothing$, $\varnothing$) & 
 ($\varnothing$, $\varnothing$) &
 ($\varnothing$, $\varnothing$)  \\
  \textbf{3123} \Tstrut\Bstrut & 
  (4.8$^{\,\Diamond}$, 11$^{\,\Diamond}$) & 
 ($\varnothing$, $\varnothing$) & 
 ($\varnothing$, $\varnothing$) &
 ($\varnothing$, $\varnothing$)  \\
\textbf{3211} \Tstrut\Bstrut & 
  ($\varnothing$, $\varnothing$) & 
  ($\varnothing$, 0.23$^{\,\Box}$) & 
  ($\varnothing$, 5.5$^{\,\Diamond}$) & 
  ($\varnothing$, 2.8$^{\,\Box}$)  \\
\textbf{3212} \Tstrut\Bstrut & 
  (4.8$^{\,\Diamond}$, 2.1$^{\,\Diamond}$) 10$^{-1}$ & 
  (2.9$^{\,\Box}$, 6.6$^{\,\Box}$) 10$^{-3}$ &
 (16$^{\,\Diamond}$, 6.6$^{\,\Diamond}$) 10$^{-2}$ &
 (1.8$^{\,\Box}$, 4.0$^{\,\Box}$) 10$^{-2}$  \\
 \textbf{3213} \Tstrut\Bstrut & 
 (3.6$^{\,\Diamond}$, 1.0$^{\,\Diamond}$) 10$^{-2}$ &  
 ($\varnothing$, $\varnothing$) & 
 (4.3$^{\,\Diamond}$, 0.6$^{\,\Diamond}$) 10$^{-2}$ & 
 ($\varnothing$, $\varnothing$) \\
\textbf{3221} \Tstrut\Bstrut & 
 ($\varnothing$, 2.2$^{\,\Box}$) &  
 (31$^{\,\Box}$, 0.54$^{\,\Box}$) 10$^{-1}$ & 
 ($\varnothing$, $\varnothing$) &  
 ($\varnothing$, 0.64$^{\,\Box}$) 10$^{-1}$ \\
\textbf{3222} \Tstrut\Bstrut & 
 (2.1$^{\,\Diamond}$, 0.85$^{\,\Diamond}$) &  
 (1.3$^{\,\Box}$, 2.7$^{\,\Box}$) 10$^{-2}$ &  
 (7.1$^{\,\Diamond}$, 2.9$^{\,\Diamond}$) 10$^{-1}$ & 
 (0.80$^{\,\Box}$, 2.4$^{\,\Box}$) 10$^{-1}$ \\
 \textbf{3223} \Tstrut\Bstrut & 
 (15$^{\,\Diamond}$, 4.8$^{\,\Diamond}$) 10$^{-2}$ &  
($\varnothing$, $\varnothing$) &
(1.9$^{\,\Diamond}$, 0.28$^{\,\Diamond}$) 10$^{-1}$ & 
($\varnothing$, $\varnothing$) \\
\textbf{3231} \Tstrut\Bstrut & 
(4.6$^{\,\triangle}$, 3.5$^{\,\triangle}$) 10$^{-1}$ &  
(4.6$^{\,\triangle}$, 3.5$^{\,\triangle}$) 10$^{-1}$ &
(2.8$^{\,\triangle}$, 2.7$^{\,\triangle}$) & 
(2.8$^{\,\triangle}$, 2.7$^{\,\triangle}$ \\
\textbf{3232} \Tstrut\Bstrut & 
(3.1$^{\,\bigtriangledown}$, 1.1$^{\,\bigtriangledown}$) &  
(0.3$^{\,\Box}$, 1.1$^{\,\bigtriangledown}$) & 
($\varnothing$, 5.8$^{\,\bigtriangledown}$) & 
(3.6$^{\,\Box}$, 5.8$^{\,\Box}$) \\
 \textbf{3233} \Tstrut\Bstrut & 
(4.0$^{\,\Diamond}$, 5.9$^{\,\Diamond}$) 10$^{-1}$ &  
 ($\varnothing$, $\varnothing$) & 
 (2.4$^{\,\Diamond}$, 3.1$^{\,\Diamond}$) & 
 ($\varnothing$, $\varnothing$)\\
\textbf{3311} \Tstrut\Bstrut & 
($\varnothing$, $\varnothing$) &  
 (5.5$^{\,\Box}$, 2.3$^{\,\Box}$) 10$^{-2}$ & 
  ($\varnothing$, $\varnothing$) & 
(6.6$^{\,\Box}$, 2.8$^{\,\Box}$) 10$^{-1}$ \\
\textbf{3312} \Tstrut\Bstrut & 
($\varnothing$, $\varnothing$) &  
 (12$^{\,\Box}$, 0.26$^{\,\Box}$) 10$^{-5}$ & 
($\varnothing$, $\varnothing$) & 
  (1.5$^{\,\Box}$, 3.5$^{\,\Box}$) 10$^{-3}$ \\
\textbf{3313} \Tstrut\Bstrut & 
 ($\varnothing$, $\varnothing$) & 
 (4.3$^{\,\triangle}$, 5.7$^{\,\triangle}$) 10$^{-2}$ & 
 ($\varnothing$, $\varnothing$) & 
 (5.2$^{\,\triangle}$, 5.7$^{\,\triangle}$) 10$^{-1}$  \\
\textbf{3321} \Tstrut\Bstrut & 
 (5.3$^{\,\Box}$, 2.3$^{\,\Box}$) 10$^{-1}$ &  
 (13$^{\,\Box}$, 5.4$^{\,\Box}$) 10$^{-3}$ &  
 (6.4$^{\,\Box}$, 2.7$^{\,\Box}$) &
 (15$^{\,\Box}$, 6.2$^{\,\Box}$) 10$^{-2}$\\
\textbf{3322} \Tstrut\Bstrut & 
 ($\varnothing$, $\varnothing$) &  
 (5.2$^{\,\Box}$, 0.11$^{\,\Box}$) 10$^{-4}$ & 
 ($\varnothing$, $\varnothing$) & 
 (0.62$^{\,\Box}$, 1.3$^{\,\Box}$) 10$^{-2}$ \\
\textbf{3323} \Tstrut\Bstrut & 
 ($\varnothing$, $\varnothing$) &  
 (1.3$^{\,\bigtriangledown}$, 0.44$^{\,\bigtriangledown}$) 10$^{-1}$ & 
 ($\varnothing$, $\varnothing$) & 
 (1.2$^{\,\bigtriangledown}$, 0.41$^{\,\bigtriangledown}$) \\
\textbf{3331} \Tstrut\Bstrut & 
 (4.7$^{\,\triangle}$, 3.5$^{\,\triangle}$) 10$^{-2}$ & 
 (4.7$^{\,\triangle}$, 3.5$^{\,\triangle}$) 10$^{-2}$ &  
 (2.8$^{\,\triangle}$, 2.7$^{\,\triangle}$) 10$^{-1}$ & 
 (2.8$^{\,\triangle}$, 2.7$^{\,\triangle}$) 10$^{-1}$ \\
\textbf{3332} \Tstrut\Bstrut & 
(2.6$^{\,\bigtriangledown}$, 0.9$^{\,\bigtriangledown}$) 10$^{-1}$ & 
(7.1$^{\,\Box}$, 2.7$^{\,\Box}$) 10$^{-4}$ & 
(1.4$^{\,\bigtriangledown}$, 0.49$^{\,\bigtriangledown}$)  &
(3.8$^{\,\Box}$, 1.4$^{\,\Box}$) 10$^{-3}$ \\
\textbf{3333} \Tstrut\Bstrut & 
 ($\varnothing$, $\varnothing$) & 
(2.4$^{\,\bigtriangledown}$, 0.83$^{\,\bigtriangledown}$) & 
 ($\varnothing$, $\varnothing$) &
(10$^{\,\bigtriangledown}$, 3.6$^{\,\bigtriangledown}$) \\ [1.1ex]
\hline
\end{tabular}}
 \caption{Same as table \ref{tab:HQ} for $C^{QuQd^{(1)}}_{ijkl}$ and $C^{QuQd^{(8)}}_{ijkl}$.}
 \label{tab:QuQd}
 \vspace{-10mm}
\end{table}
For a generic model, constraints get much more severe due not only to
the absence of the SM CKM and GIM suppression, but also to the
possible presence of right-handed flavour changing neutral currents,
which are both enhanced by RG evolution and by hadronic matrix
elements
\cite{Beall:1981ze,Bagger:1997gg,Ciuchini:1997bw,Buras:2000if}. 
One expects to constrain the ratio of the NP
coefficients over the NP scale $\Lambda$ as follows:
\begin{equation}
\label{eq:df2gen}
C^{\mathrm{NP}}_{\Delta F=2}/\Lambda^{2} <  \epsilon_{\Delta F=2} \, C^{\mathrm{SM}}_{\Delta F
  = 2}/\Lambda^{2}_{0} \ , 
\end{equation}
where for a short-distance-dominated meson mixing process 
$C^{\mathrm{SM}}_{\Delta F = 2} \sim (\lambda_\mathrm{FC}^{(d)})_{ij}^2$,
while $\epsilon_{\Delta F =i }$ is the experimentally allowed fraction of
NP contributions to the $\Delta F =i $ process times the ratio of SM over
NP matrix elements. For instance, $\Delta S = 2$ operators with
mixed chirality yield $\epsilon_{\Delta F=2} < 10^{-2}$, and plugging
in eq.~\eqref{eq:df2gen}
$(\lambda_\mathrm{FC}^{(d)})_{12} \sim 10^{-4}$ and
$C_{\Delta S=2} \sim 1$, a bound of $\Lambda \gtrsim 10^5$ TeV can be
obtained from CP violation measurements in the kaon system, see
\textit{e.g.}~\cite{Bona:2017gut}. We can now estimate the importance of the running from $\Lambda$ to $M_W$ that turns a $\Delta F=1$ operator
into a $\Delta F=2$ one. We introduce the RG
factor $\mathcal{R} \equiv \log \left(\Lambda/M_W\right)/(16 \pi^2)$,
that is order of percent for NP scales above the TeV. Inspired by
eq.~\eqref{eq:df2gen}, we naively estimate the lower limit on
$\Lambda$ from the mixing into $\Delta F=2$ to be of order
\begin{equation}
\label{eq:df1mix}
\Lambda^2 \gtrsim C_{\Delta F=1}^{\mathrm{NP}}   \frac{\mathcal{R} \,
  \Lambda_{0}^2}{\epsilon_{\Delta F=2}
  \left(\lambda_{\mathrm{FC}}^{(d),(u)}\right)_{ij}}\,.
\end{equation}

From eq.~\eqref{eq:df1mix} and the bound one analogously estimates for
$\Delta F=1$ transitions, it follows that $\Delta F=2$ constraints
overcome $\Delta F=1$ ones if
$\mathcal{R} > \epsilon_{\Delta F=2}/\epsilon_{\Delta F=1}$. This is
expected to be the case for a large class of operators, since
$\epsilon_{\Delta F=2}$ is typically at the percent level, in
particular for limits regarding CP violation, while
$\epsilon_{\Delta F=1}$ can easily be of $\mathcal{O}(1)$ or even
(much) larger for two reasons: \textit{i)} in the SM, the MFV-type
top contribution to many $\Delta F=1$ processes is not dominant with
respect to charm or light quarks; \textit{ii)} the calculation of the
relevant matrix elements turns out to be much more uncertain, often
plagued by long-distance contributions.
\begin{table}[ht!]
\resizebox{\columnwidth}{!}{
\begin{tabular}{|cV{3.0}c|c|c|}
\hline
\multirow{3}{*}{ } & \multicolumn{2}{c|}{ {\boldmath$C^{QQ^{(1,3)}}_{ijkl}$} [TeV$^{-2}$]} \Tstrut\Bstrut & \multicolumn{1}{c|}{{\boldmath$C^{dd}_{ijkl}$} [TeV$^{-2}$]} \\ [1.01ex] \clineB{2-4}{-0.1}{\boldmath$ijkl$} & $Y_{D}$ diag & $Y_{U}$ diag & $Y_{D,U}$ diag \\ [1.01ex]
\hline
\textbf{1111} \Tstrut\Bstrut & 
  5.8$^{\,\Diamond}$ 10$^{-6}$ &  
  1.4$^{\,\Box}$ 10$^{-5}$ &  
  $\varnothing$ \\
\textbf{1112} \Tstrut\Bstrut & 
 (7.0$^{\,\Diamond}$, 0.19$^{\,\Diamond}$) 10$^{-7}$ &  
 (17$^{\,\Box}$, 0.051$^{\,\Box}$) 10$^{-7}$ & 
 (3.2$^{\,\Box}$, 1.3$^{\,\Box}$) 10$^{-3}$ \\
\textbf{1113} \Tstrut\Bstrut & 
  (15$^{\,\Diamond}$, 0.44$^{\,\Diamond}$) 10$^{-6}$ &  
  (39$^{\,\triangle}$, 0.12$^{\,\Box}$) 10$^{-6}$ & 
 (1.4$^{\,\triangle}$, 1.2$^{\,\triangle}$) \\
\textbf{1122} \Tstrut\Bstrut & 
  2.9$^{\,\Diamond}$ 10$^{-6}$ &  
  6.8$^{\,\Box}$ 10$^{-6}$ &  
  $\varnothing$ \\
\textbf{1123} \Tstrut\Bstrut & 
  (5.6$^{\,\Diamond}$, 2.3$^{\,\Diamond}$) 10$^{-6}$ &  
  (1.5$^{\,\Box}$, 3.4$^{\,\Box}$) 10$^{-6}$ &  
  ($\varnothing$, $\varnothing$) \\
\textbf{1133} \Tstrut\Bstrut & 
  1.3$^{\,\Diamond}$ 10$^{-4}$ &  
  3.6$^{\,\Box}$ 10$^{-5}$ &  
  $\varnothing$ \\
\textbf{1212} \Tstrut\Bstrut & 
  (31$^{\,\Diamond}$,  0.22$^{\,\Box}$) 10$^{-8}$ &  
  (28$^{\,\Diamond}$, 0.25$^{\,\Box}$) 10$^{-8}$ &
  (65$^{\,\Box}$, 0.22$^{\,\Box}$) 10$^{-8}$ \\
\textbf{1213} \Tstrut\Bstrut & 
  (3.5$^{\,\Diamond}$, 0.1$^{\,\Diamond}$) 10$^{-6}$ &  
  (15$^{\,\Box}$, 0.3$^{\,\Box}$) 10$^{-7}$ & 
  (16$^{\,\Box}$, 0.3$^{\,\Box}$) 10$^{-4}$ \\
\textbf{1221} \Tstrut\Bstrut & 
  2.9$^{\,\Diamond}$ 10$^{-6}$ &  
  6.8$^{\,\Box}$ 10$^{-6}$&  
  $\varnothing$ \\
\textbf{1222} \Tstrut\Bstrut & 
  (7.0$^{\,\Diamond}$, 0.19$^{\,\Diamond}$) 10$^{-7}$ &  
  (17$^{\,\Box}$, 0.05$^{\,\Box}$) 10$^{-7}$ &
  (3.2$^{\,\Box}$, 1.3$^{\,\Box}$) 10$^{-3}$ \\
\textbf{1223} \Tstrut\Bstrut & 
  (15$^{\,\Diamond}$, 0.44$^{\,\Diamond}$) 10$^{-6}$ &  
  (39$^{\,\Diamond}$, 0.12$^{\,\Diamond}$) 10$^{-6}$ &  
  ($\varnothing$, $\varnothing$) \\
\textbf{1231} \Tstrut\Bstrut & 
  (5.6$^{\,\Diamond}$, 2.3$^{\,\Diamond}$) 10$^{-6}$ &  
  (1.5$^{\,\Box}$, 3.4$^{\,\Box}$) 10$^{-6}$ &   
  ($\varnothing$, $\varnothing$) \\
\textbf{1232} \Tstrut\Bstrut & 
  (1.3$^{\,\Diamond}$, 2.2$^{\,\Diamond}$) 10$^{-6}$ &  
  (3.7$^{\,\Box}$, 1.4$^{\,\Box}$) 10$^{-7}$ &
  (7.2$^{\,\Box}$, 2.9$^{\,\Box}$) 10$^{-3}$ \\
\textbf{1233} \Tstrut\Bstrut & 
  (3.1$^{\,\Diamond}$, 6.2$^{\,\Diamond}$) 10$^{-5}$ & 
  (8.9$^{\,\Box}$, 3.4$^{\,\Box}$) 10$^{-6}$ &  
  ($\varnothing$, $\varnothing$) \\
\textbf{1313} \Tstrut\Bstrut & 
  (1.1$^{\,\triangle}$, 0.90$^{\,\triangle}$) 10$^{-6}$ &  
  (1.1$^{\,\triangle}$, 0.95$^{\,\triangle}$) 10$^{-6}$ &
  (1.0$^{\,\triangle}$, 0.88$^{\,\triangle}$) 10$^{-6}$ \\
\textbf{1322} \Tstrut\Bstrut & 
  (15$^{\,\Diamond}$, 0.44$^{\,\Diamond}$) 10$^{-6}$ &   
  (39$^{\,\Diamond}$, 0.12$^{\,\Diamond}$) 10$^{-6}$ &  
  ($\varnothing$, $\varnothing$) \\
\textbf{1323} \Tstrut\Bstrut & 
  (3.3$^{\,\Diamond}$, 0.1$^{\,\Diamond}$) 10$^{-4}$ &  
  (2.4$^{\,\triangle}$, 2.1$^{\,\triangle}$) 10$^{-6}$ &
  (6.3$^{\,\triangle}$, 5.3$^{\,\triangle}$) 10$^{-1}$ \\
\textbf{1331} \Tstrut\Bstrut & 
  1.3$^{\,\Diamond}$ 10$^{-4}$ & 
  3.6$^{\,\Box}$ 10$^{-5}$ &  
  $\varnothing$ \\
\textbf{1332} \Tstrut\Bstrut & 
  (3.1$^{\,\Diamond}$, 6.2$^{\,\Diamond}$) 10$^{-5}$ &   
  (8.9$^{\,\Box}$, 3.4$^{\,\Box}$) 10$^{-6}$ &  
  ($\varnothing$, $\varnothing$) \\
\textbf{1333} \Tstrut\Bstrut & 
  (6.7$^{\,\Diamond}$, 15$^{\,\Diamond}$) 10$^{-4}$ &  
  (6.9$^{\,\triangle}$, 5.8$^{\,\triangle}$) 10$^{-5}$ &
  (1.4$^{\,\triangle}$, 1.2$^{\,\triangle}$)\\
\textbf{2222} \Tstrut\Bstrut & 
  5.8$^{\,\Diamond}$ 10$^{-6}$ &  
  1.4$^{\,\Box}$ 10$^{-5}$ &  
  $\varnothing$ \\
\textbf{2223} \Tstrut\Bstrut & 
  (5.7$^{\,\Diamond}$, 2.3$^{\,\Diamond}$) 10$^{-6}$ &  
  (1.5$^{\,\Box}$, 3.4$^{\,\Box}$) 10$^{-6}$ &
  (3.0$^{\,\bigtriangledown}$, 0.98$^{\,\bigtriangledown}$) 10$^{-1}$ \\
\textbf{2233} \Tstrut\Bstrut & 
  1.3$^{\,\Diamond}$ 10$^{-4}$ &  
  3.6$^{\,\Box}$ 10$^{-5}$ &  
  $\varnothing$ \\
\textbf{2323} \Tstrut\Bstrut & 
  (2.2$^{\,\bigtriangledown}$, 0.75$^{\,\bigtriangledown}$) 10$^{-5}$ &  
  (2.1$^{\,\triangle}$, 0.80$^{\,\triangle}$) 10$^{-5}$ &
  (2.2$^{\,\bigtriangledown}$, 0.74$^{\,\bigtriangledown}$) 10$^{-5}$ \\
\textbf{2332} \Tstrut\Bstrut & 
  1.3$^{\,\Diamond}$ 10$^{-4}$ &  
  3.6$^{\,\Box}$ 10$^{-5}$ &  
  $\varnothing$ \\
\textbf{2333} \Tstrut\Bstrut & 
  (2.2$^{\,\Diamond}$, 2.8$^{\,\Diamond}$) 10$^{-3}$ &  
  (2.8$^{\,\bigtriangledown}$, 0.94$^{\,\bigtriangledown}$) 10$^{-4}$ &
  (3.0$^{\,\bigtriangledown}$, 0.98$^{\,\bigtriangledown}$) 10$^{-1}$ \\
\textbf{3333} \Tstrut\Bstrut & 
  4.2$^{\,\Diamond}$ 10$^{-1}$ & 
  1.3$^{\,\bigtriangledown}$ 10$^{-2}$ &
  $ \varnothing $ \\[1.1ex]
\hline
\end{tabular}}
 \caption{Same as table \ref{tab:HQ} for $C^{QQ^{(1,3)}}_{ijkl}$ and $C^{dd}_{ijkl}$.}
 \label{tab:QQuudd}
 \vspace{-3mm}
\end{table} 
Having argued on the general relevance of $\Delta F=2$ constraints, we
present in the following the results of our study and compare them to
a few examples from the literature. We
collect our main findings in tables~\ref{tab:HQ}-\ref{tab:QdQu}. A
more general study accounting for all other bounds coming from
$\Delta F=1$ measurements and EW precision tests will be given
elsewhere. Previous work in this direction can be found in~\cite{Endo:2016tnu,Bobeth:2017xry,Feruglio:2017rjo,Gonzalez-Alonso:2017iyc,Buttazzo:2017ixm,Kumar:2018kmr,Aebischer:2018iyb}.

\begin{table*}[ht!]
\resizebox{\textwidth}{!}{
\begin{tabular}{|cV{3.0}c|c|c|c|c|c|c|c|c|}
\hline
\multirow{3}{*}{ }& \multicolumn{2}{c|}{ {\boldmath$C^{Qd^{(1)}}_{ijkl}$} [TeV$^{-2}$] } \Tstrut\Bstrut  & 
\multicolumn{2}{c|}{ {\boldmath$C^{Qd^{(8)}}_{ijkl}$} [TeV$^{-2}$] } & \multicolumn{2}{c|}{ {\boldmath$C^{Qu^{(1)}}_{ijkl}$} [TeV$^{-2}$] } &
\multicolumn{2}{c|}{ {\boldmath$C^{Qu^{(8)}}_{ijkl}$} [TeV$^{-2}$] } \\ [1.01ex]
 \clineB{2-9}{-0.1}
 {\boldmath${ijkl}$} & $Y_{D}$ diag & $Y_{U}$ diag & $Y_{D}$ diag & $Y_{U}$ diag & $Y_{D}$ diag & $Y_{U}$ diag & $Y_{D}$ diag & $Y_{U}$ diag \\ [1.01ex]
\hline
\textbf{1111} \Tstrut\Bstrut & 
 11$^{\,\Box}$  &  
 8.2$^{\,\Box}$ 10$^{-4}$ &  
 3.3$^{\,\Box}$ & 
 1.7$^{\,\Box}$ 10$^{-4}$& 
 3.7$^{\,\Diamond}$ 10$^{-1}$&  
 $\varnothing$ & 
 2.2$^{\,\Diamond}$  & 
 $\varnothing$ \\
\textbf{1112} \Tstrut\Bstrut & 
 (1.9$^{\,\Box}$, 0.81$^{\,\Box}$) 10$^{-5}$ &  
 (64$^{\,\Box}$, 0.29$^{\,\Box}$) 10$^{-9}$ & 
 (5.5$^{\,\Box}$, 2.3$^{\,\Box}$) 10$^{-6}$ & 
 (14$^{\,\Box}$, 0.062$^{\,\Box}$) 10$^{-9}$ & 
 (6.0$^{\,\Diamond}$, 0.18$^{\Diamond}$) 10$^{-7}$ &  
 (5.0$^{\,\Diamond}$, 4.9$^{\Diamond}$) & 
 (14$^{\,\Diamond}$, 0.45$^{\Diamond}$) 10$^{-8}$ & 
 (1.4$^{\,\Diamond}$, 1.4$^{\Diamond}$)  \\
\textbf{1113} \Tstrut\Bstrut & 
 (2.2$^{\,\triangle}$, 2.1$^{\,\triangle}$) 10$^{-3}$ &  
 (7.7$^{\,\triangle}$, 0.77$^{\,\Box}$) 10$^{-5}$& 
 (1.1$^{\,\triangle}$, 0.81$^{\,\triangle}$) 10$^{-3}$& 
 (3.4$^{\,\triangle}$, 0.16$^{\,\Box}$) 10$^{-5}$&
 (3.9$^{\,\Diamond}$, 0.12$^{\,\Diamond}$) 10$^{-2}$ &  
 ($\varnothing$, 1.8$^{\,\Box}$) & 
 (24$^{\,\Diamond}$, 0.74$^{\,\Diamond}$) 10$^{-2}$ & 
 ($\varnothing$, 5.5$^{\,\Box}$) \\
\textbf{1122} \Tstrut\Bstrut & 
 12$^{\,\Box}$ &  
 8.2$^{\,\Box}$ 10$^{-4}$ & 
 3.3$^{\,\Box}$ & 
 1.7$^{\,\Box}$ 10$^{-4}$ & 
 1.9$^{\,\Diamond}$ 10$^{-1}$ &  
 $\varnothing$ & 
 1.4$^{\,\Diamond}$ 10$^{-1}$ & 
 $\varnothing$ \\
 \textbf{1123} \Tstrut\Bstrut & 
 ($\varnothing$,  $\varnothing$) &  
 (1.9$^{\,\Box}$, 0.75$^{\,\Box}$) 10$^{-3}$ & 
 (7.4$^{\,\Box}$, 7.4$^{\,\Box}$) & 
 (4.0$^{\,\Box}$, 1.6$^{\,\Box}$) 10$^{-4}$ & 
 (7.9$^{\,\Diamond}$, 6.5$^{\,\Diamond}$) 10$^{-3}$ &  
 (4.8$^{\,\Box}$, 11$^{\,\Box}$) 10$^{-2}$& 
 (5.9$^{\,\Diamond}$, 4.9$^{\,\Diamond}$) 10$^{-3}$ & 
 (1.4$^{\,\Box}$, 3.2$^{\,\Box}$) 10$^{-1}$ \\
\textbf{1133}  \Tstrut\Bstrut & 
 $\varnothing$ &  
 $\varnothing$ & 
 $\varnothing$ & 
 10$^{\,\Box}$ & 
 $\varnothing$ &  
 4.1$^{\,\Box}$ 10$^{-3}$ & 
 $\varnothing$ & 
 1.2$^{\,\Box}$ 10$^{-2}$ \\
\textbf{1211} \Tstrut\Bstrut & 
 (1.8$^{\,\Box}$, 0.76$^{\,\Box}$) 10$^{-4}$ &  
 (2.0$^{\,\Box}$, 0.76$^{\,\Box}$) 10$^{-4}$ & 
 (3.8$^{\,\Box}$, 1.6$^{\,\Box}$) 10$^{-5}$ & 
 (4.3$^{\,\Box}$, 1.6$^{\,\Box}$) 10$^{-5}$ &
 (9.1$^{\,\Diamond}$, 2.5$^{\,\Diamond}$) 10$^{-2}$ &  
 ($\varnothing$, $\varnothing$) & 
 (5.4$^{\,\Diamond}$, 1.5$^{\,\Diamond}$) 10$^{-1}$& 
 ($\varnothing$, $\varnothing$) \\
\textbf{1212} \Tstrut\Bstrut & 
 (140$^{\,\Box}$, 0.63$^{\,\Box}$) 10$^{-10}$ &  
 (150$^{\,\Box}$, 0.67$^{\,\Box}$) 10$^{-10}$ & 
 (30$^{\,\Box}$, 0.14$^{\,\Box}$) 10$^{-10}$ & 
 (32$^{\,\Box}$, 0.14$^{\,\Box}$) 10$^{-10}$ & 
 (14$^{\,\Diamond}$, 0.43$^{\,\Diamond}$) 10$^{-8}$&  
 (13$^{\,\Diamond}$, 0.40$^{\,\Diamond}$) 10$^{-8}$&
 (3.3$^{\,\Diamond}$, 0.1$^{\,\Diamond}$) 10$^{-8}$& 
 (32$^{\,\Diamond}$, 0.98$^{\,\Diamond}$) 10$^{-9}$ \\
\textbf{1213} \Tstrut\Bstrut & 
 (9.2$^{\,\Box}$, 0.17$^{\,\Box}$) 10$^{-5}$ &  
 (6.4$^{\,\triangle}$, 1.8$^{\,\Box}$) 10$^{-6}$ & 
 (20$^{\,\Box}$, 0.36$^{\,\Box}$) 10$^{-6}$ & 
 (36$^{\,\triangle}$, 3.8$^{\,\triangle}$) 10$^{-7}$ & 
 (9.1$^{\,\Diamond}$, 0.28$^{\,\Diamond}$) 10$^{-3}$ &  
 ($\varnothing$, 4.4$^{\,\Diamond}$) 10$^{-1}$ & 
 (5.5$^{\,\Diamond}$, 0.17$^{\,\Diamond}$) 10$^{-2}$ & 
  ($\varnothing$, 5.2$^{\,\Diamond}$) 10$^{-1}$    \\
\textbf{1221} \Tstrut\Bstrut & 
 (1.2$^{\,\Box}$, 1.2$^{\,\Box}$) &  
 (28$^{\,\Box}$, 0.13$^{\,\Box}$) 10$^{-8}$ & 
 (2.5$^{\,\Box}$, 2.7$^{\,\Box}$) 10$^{-1}$ &  
 (60$^{\,\Box}$, 0.27$^{\,\Box}$) 10$^{-9}$ &  
 (26$^{\,\Diamond}$, 0.8$^{\,\Box}$) 10$^{-7}$ &
 ($\varnothing$, $\varnothing$) &
 (6.2$^{\,\Diamond}$, 0.19$^{\,\Diamond}$) 10$^{-7}$ & 
 ($\varnothing$, $\varnothing$) \\
\textbf{1222} \Tstrut\Bstrut & 
 (1.8$^{\,\Box}$, 0.76$^{\,\Box}$) 10$^{-4}$ &  
 (2.0$^{\,\Box}$, 0.77$^{\,\Box}$) 10$^{-4}$ & 
 (3.9$^{\,\Box}$, 1.6$^{\,\Box}$) 10$^{-5}$ & 
 (4.3$^{\,\Box}$, 1.6$^{\,\Box}$) 10$^{-5}$ & 
 (4.5$^{\,\Diamond}$, 1.3$^{\,\Diamond}$) 10$^{-2}$ & 
 ($\varnothing$, 4.5$^{\,\Diamond}$) 10$^{-2}$ &
 (3.4$^{\,\Diamond}$, 0.95$^{\,\Diamond}$) 10$^{-2}$ &  
 ($\varnothing$, 1.3$^{\,\Diamond}$) 10$^{-1}$ \\
\textbf{1223} \Tstrut\Bstrut & 
 (4.6$^{\,\Box}$, 2.1$^{\,\Box}$) &  
 (7.3$^{\,\bigtriangledown}$, 2.5$^{\,\bigtriangledown}$) 10$^{-4}$ & 
 (9.8$^{\,\Box}$, 4.4$^{\,\Box}$) 10$^{-1}$ & 
 (4.6$^{\,\bigtriangledown}$, 1.6$^{\,\bigtriangledown}$) 10$^{-4}$ & 
 (3.4$^{\,\Diamond}$, 2.8$^{\,\Diamond}$) 10$^{-2}$ &  
 (120$^{\,\Box}$, 0.38$^{\,\Box}$) 10$^{-2}$ & 
 (2.6$^{\,\Diamond}$, 2.1$^{\,\Diamond}$) 10$^{-2}$ & 
 (37$^{\,\Box}$, 0.11$^{\,\Box}$) 10$^{-1}$ \\
\textbf{1231} \Tstrut\Bstrut & 
 ($\varnothing$, $\varnothing$) &  
 (3.3$^{\,\triangle}$, 0.33$^{\,\Box}$) 10$^{-4}$ & 
 ($\varnothing$, $\varnothing$) & 
 (15$^{\,\triangle}$, 0.71$^{\,\Box}$) 10$^{-5}$ & 
 (4.8$^{\,\Box}$, 0.53$^{\,\Diamond}$) 10$^{-2}$ &  
 ($\varnothing$, $\varnothing$) & 
 (14$^{\,\Box}$, 3.2$^{\,\Diamond}$) 10$^{-2}$ & 
 ($\varnothing$, $\varnothing$) \\
\textbf{1232} \Tstrut\Bstrut & 
 (4.1$^{\,\Box}$, 1.7$^{\,\Box}$) 10$^{-4}$ &  
 (4.3$^{\,\Box}$, 1.7$^{\,\Box}$) 10$^{-4}$& 
 (8.7$^{\,\Box}$, 3.6$^{\,\Box}$) 10$^{-5}$ & 
 (9.2$^{\,\Box}$, 3.7$^{\,\Box}$) 10$^{-5}$ &  
 (3.8$^{\,\Diamond}$, 1.1$^{\,\Diamond}$) 10$^{-3}$ &  
 (12$^{\,\Box}$, 4.4$^{\,\Box}$) 10$^{-3}$ & 
 (2.8$^{\,\Diamond}$, 0.82$^{\,\Diamond}$) 10$^{-3}$ &  
 (3.5$^{\,\Box}$, 1.3$^{\,\Box}$) 10$^{-2}$ \\
\textbf{1233} \Tstrut\Bstrut & 
 (10$^{\,\Box}$, 4.5$^{\,\Box}$) &  
 (12$^{\,\Box}$, 4.3$^{\,\Box}$) & 
 (2.2$^{\,\Box}$, 0.95$^{\,\Box}$) & 
 (2.5$^{\,\Box}$, 0.95$^{\,\Box}$) & 
 (2.5$^{\,\Box}$, 1.0$^{\,\Box}$) 10$^{-1}$ &  
 (9.9$^{\,\Box}$, 3.8$^{\,\Box}$) 10$^{-4}$ & 
 (7.6$^{\,\Box}$, 3.1$^{\,\Box}$) 10$^{-1}$ &  
 (3.0$^{\,\Box}$, 1.1$^{\,\Box}$) 10$^{-3}$ \\
\textbf{1311} \Tstrut\Bstrut & 
 (1.8$^{\,\Box}$, 0.80$^{\,\Box}$) 10$^{-1}$ &  
 (4.7$^{\,\Box}$, 1.8$^{\,\Box}$) 10$^{-3}$ & 
 (3.0$^{\,\Box}$, 1.4$^{\,\Box}$) 10$^{-2}$ &  
 (10$^{\,\Box}$, 4.0$^{\,\Box}$) 10$^{-4}$ & 
 (1.8$^{\,\Diamond}$, 0.6$^{\,\Diamond}$) &  
 ($\varnothing$, $\varnothing$) & 
 (11$^{\,\Diamond}$, 3.5$^{\,\Diamond}$) & 
 ($\varnothing$, $\varnothing$) \\
\textbf{1312} \Tstrut\Bstrut & 
 (36$^{\,\Box}$, 0.65$^{\,\Box}$) 10$^{-7}$ &  
 (8.6$^{\,\Box}$, 0.16$^{\,\Box}$) 10$^{-8}$ & 
 (6.1$^{\,\Box}$, 0.11$^{\,\Box}$) 10$^{-7}$ & 
 (18$^{\,\Box}$, 0.34$^{\,\Box}$) 10$^{-9}$ & 
 (31$^{\,\Diamond}$, 0.96$^{\,\Diamond}$) 10$^{-7}$ &  
 (3.6$^{\,\Diamond}$, 0.11$^{\,\Diamond}$) 10$^{-1}$ & 
 (7.5$^{\,\Diamond}$, 0.23$^{\,\Diamond}$) 10$^{-7}$ &  
 (9.4$^{\,\Diamond}$, 0.29$^{\,\Diamond}$) 10$^{-2}$ \\
\textbf{1313} \Tstrut\Bstrut & 
 (2.6$^{\,\triangle}$, 2.5$^{\,\triangle}$) 10$^{-7}$ &  
 (2.7$^{\,\triangle}$, 2.6$^{\,\triangle}$) 10$^{-7}$ & 
 (1.5$^{\,\triangle}$, 1.1$^{\,\triangle}$) 10$^{-7}$ & 
 (1.6$^{\,\triangle}$, 1.2$^{\,\triangle}$) 10$^{-7}$ & 
 (20$^{\,\Diamond}$, 0.63$^{\,\Diamond}$) 10$^{-2}$ &  
 (8.5$^{\,\triangle}$, 7.1$^{\,\triangle}$) & 
 (12$^{\,\Diamond}$, 0.38$^{\,\Diamond}$) 10$^{-1}$ & 
 ($\varnothing$, $\varnothing$) \\
\textbf{1321} \Tstrut\Bstrut & 
 (5.2$^{\,\Box}$, 5.2$^{\,\Box}$) 10$^{-2}$ &  
 (8.6$^{\,\Box}$, 3.5$^{\,\Box}$) 10$^{-8}$ & 
 (1.1$^{\,\Box}$, 1.1$^{\,\Box}$) 10$^{-2}$ & 
 (1.8$^{\,\Box}$, 0.75$^{\,\Box}$) 10$^{-8}$ & 
 (5.2$^{\,\Diamond}$, 12$^{\,\Diamond}$) 10$^{-6}$ &  
 ($\varnothing$, $\varnothing$) & 
 (1.3$^{\,\Diamond}$, 2.8$^{\,\Diamond}$) 10$^{-6}$ & 
 ($\varnothing$, $\varnothing$) \\
\textbf{1322} \Tstrut\Bstrut & 
 (1.8$^{\,\Box}$, 0.8$^{\,\Box}$) 10$^{-1}$ &  
 (4.7$^{\,\Box}$, 1.8$^{\,\Box}$) 10$^{-3}$ & 
 (3.0$^{\,\Box}$, 1.4$^{\,\Box}$) 10$^{-2}$ & 
 (10$^{\,\Box}$, 4.0$^{\,\Box}$) 10$^{-4}$ & 
 (9.2$^{\,\Diamond}$, 3.0$^{\,\Diamond}$) 10$^{-1}$ &  
 ($\varnothing$, 1.1$^{\,\Box}$) & 
 (7.0$^{\,\Diamond}$, 2.2$^{\,\Diamond}$) 10$^{-1}$ & 
 ($\varnothing$, 3.2$^{\,\Box}$)  \\
\textbf{1323} \Tstrut\Bstrut & 
 (3.2$^{\,\triangle}$, 3.1$^{\,\triangle}$) 10$^{-1}$ &  
 (3.1$^{\,\bigtriangledown}$, 1.1$^{\,\bigtriangledown}$) 10$^{-5}$ & 
 (1.8$^{\,\triangle}$, 1.4$^{\,\triangle}$) 10$^{-1}$ &  
 (2.0$^{\,\bigtriangledown}$, 0.67$^{\,\bigtriangledown}$) 10$^{-5}$ & 
 (7.4$^{\,\triangle}$, 6.2$^{\,\triangle}$) 10$^{-2}$ &  
 (7.6$^{\,\triangle}$, 6.4$^{\,\triangle}$) 10$^{-2}$ & 
 (2.2$^{\,\triangle}$, 1.9$^{\,\triangle}$) 10$^{-1}$ &  
 (2.3$^{\,\triangle}$, 2.0$^{\,\triangle}$) 10$^{-1}$ \\
\textbf{1331} \Tstrut\Bstrut & 
 (8.9$^{\,\triangle}$, 8.7$^{\,\triangle}$) &  
 (2.2$^{\,\Box}$, 0.94$^{\,\Box}$) 10$^{-3}$ & 
 (4.0$^{\,\triangle}$, 4.0$^{\,\triangle}$) & 
 (4.7$^{\,\Box}$, 2.0$^{\,\Box}$) 10$^{-4}$ &  
 (3.5$^{\,\Diamond}$, 6.7$^{\,\triangle}$) 10$^{-1}$ &   
 ($\varnothing$, $\varnothing$) & 
 (2.1$^{\,\Diamond}$, 2.0$^{\,\triangle}$) &  
 ($\varnothing$, $\varnothing$) \\
\textbf{1332} \Tstrut\Bstrut & 
 (4.0$^{\,\Box}$, 2.7$^{\,\Box}$) 10$^{-1}$ & 
 (4.2$^{\,\Box}$, 0.25$^{\,\Box}$) 10$^{-3}$ & 
 (6.9$^{\,\Box}$, 3.0$^{\,\Box}$) 10$^{-2}$ & 
 (15$^{\,\Box}$, 2.2$^{\,\Box}$) 10$^{-4}$ &  
 (15$^{\,\Diamond}$, 4.7$^{\,\Diamond}$) 10$^{-5}$ &  
 (2.8$^{\,\Box}$, 1.1$^{\,\Box}$) 10$^{-1}$ & 
 (11$^{\,\Diamond}$, 3.5$^{\,\Diamond}$) 10$^{-5}$ &   
 (8.3$^{\,\Box}$, 3.2$^{\,\Box}$) 10$^{-1}$ \\
\textbf{1333} \Tstrut\Bstrut & 
 (7.0$^{\,\triangle}$, 7.0$^{\,\triangle}$) 10$^{-1}$ &  
 (7.1$^{\,\triangle}$, 2.8$^{\,\bigtriangledown}$) 10$^{-1}$ & 
 (4.0$^{\,\triangle}$, 3.0$^{\,\triangle}$) 10$^{-1}$ &   
 (3.8$^{\,\triangle}$, 1.8$^{\,\bigtriangledown}$) 10$^{-1}$ &  
 (7.4$^{\,\triangle}$, 6.3$^{\,\triangle}$) 10$^{-3}$ &    
 (7.6$^{\,\triangle}$, 6.4$^{\,\triangle}$) 10$^{-3}$ &  
 (2.2$^{\,\triangle}$, 1.9$^{\,\triangle}$) 10$^{-2}$ &  
 (2.3$^{\,\triangle}$, 2.0$^{\,\triangle}$) 10$^{-2}$ \\
\textbf{2211} \Tstrut\Bstrut & 
 11$^{\,\Box}$ &  
 8.3$^{\,\Box}$ 10$^{-4}$ & 
 2.0$^{\,\Box}$ & 
 1.8$^{\,\Box}$ 10$^{-4}$ &  
 3.7$^{\,\Diamond}$ 10$^{-1}$ &  
 $\varnothing$ & 
 2.2$^{\,\Diamond}$ &   
 $\varnothing$ \\
\textbf{2212} \Tstrut\Bstrut & 
 (1.9$^{\,\Box}$, 0.79$^{\,\Box}$) 10$^{-5}$ &   
 (64$^{\,\Box}$, 0.29$^{\,\Box}$) 10$^{-9}$ & 
 (3.2$^{\,\Box}$, 1.3$^{\,\Box}$) 10$^{-6}$ &  
 (140$^{\,\Box}$, 0.62$^{\,\Box}$) 10$^{-10}$ & 
 (6.0$^{\,\Diamond}$, 0.18$^{\,\Diamond}$) 10$^{-7}$ &  
 (3.2$^{\,\Diamond}$, 3.2$^{\,\Diamond}$) & 
 (14$^{\,\Diamond}$, 0.45$^{\,\Diamond}$) 10$^{-8}$ &  
 (8.5$^{\,\Diamond}$, 8.4$^{\,\Diamond}$) 10$^{-1}$ \\
\textbf{2213} \Tstrut\Bstrut & 
 (4.8$^{\,\Box}$, 2.1$^{\,\Box}$) 10$^{-1}$  &  
 (2.8$^{\,\triangle}$, 0.77$^{\,\Box}$) 10$^{-5}$  &  
 (8.1$^{\,\Box}$, 3.6$^{\,\Box}$) 10$^{-2}$  &   
 (1.6$^{\,\Box}$, 0.16$^{\,\Box}$) 10$^{-5}$  &  
 (4.0$^{\,\Diamond}$, 0.12$^{\,\Diamond}$) 10$^{-2}$  &  
 ($\varnothing$, 1.8$^{\,\Box}$)  & 
 (24$^{\,\Diamond}$, 0.74$^{\,\Diamond}$) 10$^{-2}$ & 
 ($\varnothing$, 5.5$^{\,\Box}$)  \\
\textbf{2222} \Tstrut\Bstrut & 
 11$^{\,\Box}$ &  
 8.2$^{\,\Box}$ 10$^{-4}$ & 
 2.0$^{\,\Box}$ & 
 1.8$^{\,\Box}$ 10$^{-4}$ &  
 1.9$^{\,\Diamond}$ 10$^{-1}$ &  
 $\varnothing$ & 
 1.4$^{\,\Diamond}$ 10$^{-1}$ &   
 $\varnothing$ \\
\textbf{2223} \Tstrut\Bstrut & 
 (1.2$^{\,\bigtriangledown}$, 0.39$^{\,\bigtriangledown}$) 10$^{-2}$ &  
 (1.7$^{\,\bigtriangledown}$, 0.57$^{\,\bigtriangledown}$) 10$^{-4}$ &  
 (6.1$^{\,\bigtriangledown}$, 2.1$^{\,\bigtriangledown}$) 10$^{-3}$ & 
 (1.1$^{\,\bigtriangledown}$, 0.36$^{\,\bigtriangledown}$) 10$^{-4}$ & 
 (16$^{\,\Diamond}$, 4.7$^{\,\Diamond}$) 10$^{-3}$ &  
 (4.8$^{\,\Box}$, 11$^{\,\Box}$) 10$^{-2}$ &  
  (12$^{\,\Diamond}$, 3.5$^{\,\Diamond}$) 10$^{-3}$ & 
 (1.4$^{\,\Box}$, 3.2$^{\,\Box}$) 10$^{-1}$ \\
\textbf{2233} \Tstrut\Bstrut & 
 $\varnothing$ &  
 4.5$^{\,\bigtriangledown}$  & 
 ${\varnothing}$ & 
 2.9$^{\,\bigtriangledown}$ &  
 ${\varnothing}$ & 
 4.1$^{\,\Box}$ 10$^{-3}$ & 
  $\varnothing$ &   
  1.2$^{\,\Box}$ 10$^{-2}$ \\
\textbf{2311} \Tstrut\Bstrut & 
 (4.9$^{\,\Box}$, 4.9$^{\,\Box}$) 10$^{-1}$ &  
 (0.72$^{\,\Box}$, 2.7$^{\,\Box}$) 10$^{-2}$ &  
 (8.3$^{\,\Box}$, 8.4$^{\,\Box}$) 10$^{-2}$ & 
 (1.5$^{\,\Box}$, 6.0$^{\,\Box}$) 10$^{-3}$ &
 ($\varnothing$, 3.5$^{\,\Diamond}$) &
 ($\varnothing$, $\varnothing$) & 
 ($\varnothing$, $\varnothing$) & 
 ($\varnothing$, $\varnothing$) \\
\textbf{2312} \Tstrut\Bstrut & 
 (1.8$^{\,\Box}$, 0.8$^{\,\Box}$) 10$^{-2}$ &  
 (39$^{\,\Box}$, 0.68$^{\,\Box}$) 10$^{-8}$ & 
 (2.6$^{\,\Box}$, 1.2$^{\,\Box}$) 10$^{-3}$ & 
 (8.3$^{\,\Box}$, 0.15$^{\,\Box}$) 10$^{-8}$ & 
 (13$^{\,\Diamond}$, 0.42$^{\,\Diamond}$) 10$^{-6}$ & 
 ($\varnothing$, $\varnothing$) & 
 (3.2$^{\,\Diamond}$, 0.10$^{\,\Diamond}$) 10$^{-6}$ & 
 ($\varnothing$, $\varnothing$) \\
\textbf{2313} \Tstrut\Bstrut & 
 (2.4$^{\,\triangle}$, 2.3$^{\,\triangle}$) 10$^{-2}$ &  
 (1.2$^{\,\triangle}$, 1.1$^{\,\triangle}$) 10$^{-6}$ &  
 (1.5$^{\,\triangle}$, 1.1$^{\,\triangle}$) 10$^{-2}$ &   
 (6.7$^{\,\triangle}$, 5.0$^{\,\triangle}$) 10$^{-7}$ &  
 (8.9$^{\,\Diamond}$, 0.27$^{\,\Diamond}$) 10$^{-1}$ &  
 ($\varnothing$, $\varnothing$) & 
 (5.3$^{\,\Diamond}$, 0.17$^{\,\Diamond}$) & 
 ($\varnothing$, $\varnothing$) \\
\textbf{2321} \Tstrut\Bstrut & 
 (8.3$^{\,\Box}$, 3.4$^{\,\Box}$) 10$^{-7}$ &  
 (2.0$^{\,\Box}$, 0.81$^{\,\Box}$) 10$^{-8}$ &  
 (14$^{\,\Box}$, 5.7$^{\,\Box}$) 10$^{-8}$ &  
 (8.3$^{\,\Box}$, 1.7$^{\,\Box}$) 10$^{-9}$ &
 (1.2$^{\,\Diamond}$, 2.7$^{\,\Diamond}$) 10$^{-6}$ &
 (1.4$^{\,\Diamond}$, 3.1$^{\,\Diamond}$) 10$^{-1}$ &
 (3.0$^{\,\Diamond}$, 6.5$^{\,\Diamond}$) 10$^{-7}$ & 
 (3.6$^{\,\Diamond}$, 8.2$^{\,\Diamond}$) 10$^{-2}$ \\
\textbf{2322} \Tstrut\Bstrut & 
 (1.9$^{\,\bigtriangledown}$, 0.63$^{\,\bigtriangledown}$) &
 (0.72$^{\,\Box}$, 2.8$^{\,\Box}$) 10$^{-2}$ &
 (8.3$^{\,\Box}$, 4.0$^{\,\bigtriangledown}$) 10$^{-2}$ &
 (1.5$^{\,\Box}$, 6.0$^{\,\Box}$) 10$^{-3}$ & 
 (8.6$^{\,\bigtriangledown}$, 1.8$^{\,\Diamond}$) &
 ($\varnothing$, $\varnothing$) & 
 ($\varnothing$, 1.3$^{\,\Diamond}$ ) & 
 ($\varnothing$, $\varnothing$) \\
\textbf{2323} \Tstrut\Bstrut & 
 (7.0$^{\,\bigtriangledown}$, 2.4$^{\,\bigtriangledown}$) 10$^{-6}$ &  
 (7.2$^{\,\bigtriangledown}$, 2.4$^{\,\bigtriangledown}$) 10$^{-6}$ &  
 (4.4$^{\,\bigtriangledown}$, 1.5$^{\,\bigtriangledown}$) 10$^{-6}$ &   
 (4.5$^{\,\bigtriangledown}$, 1.6$^{\,\bigtriangledown}$) 10$^{-6}$ & 
 (3.6$^{\,\Diamond}$, 1.2$^{\,\bigtriangledown}$) 10$^{-1}$ &  
 (3.3$^{\,\triangle}$, 1.2$^{\,\bigtriangledown}$) 10$^{-1}$ &  
 (2.7$^{\,\Diamond}$, 3.2$^{\,\bigtriangledown}$) 10$^{-1}$ &  
 (10$^{\,\triangle}$, 3.7$^{\,\bigtriangledown}$) 10$^{-1}$ \\
\textbf{2331} \Tstrut\Bstrut & 
 (2.1$^{\,\Box}$, 1.0$^{\,\Box}$) 10$^{-2}$ &  
 (5.1$^{\,\Box}$, 2.2$^{\,\Box}$) 10$^{-4}$ &
 (3.6$^{\,\Box}$, 1.5$^{\,\Box}$) 10$^{-3}$ &
 (11$^{\,\Box}$, 4.7$^{\,\Box}$) 10$^{-5}$ &
 (0.46$^{\,\Diamond}$, 1.8$^{\,\Diamond}$) 10$^{-1}$& 
 ($\varnothing$, $\varnothing$) & 
 (4.8$^{\,\Diamond}$, 11$^{\,\Diamond}$) 10$^{-1}$ & 
 ($\varnothing$, $\varnothing$) \\
\textbf{2332} \Tstrut\Bstrut & 
 (7.4$^{\,\bigtriangledown}$, 2.5$^{\,\bigtriangledown}$) &
 (4.0$^{\,\bigtriangledown}$, 1.1$^{\,\Box}$) 10$^{-3}$  &
 (4.7$^{\,\bigtriangledown}$, 1.6$^{\,\bigtriangledown}$) &
 (2.5$^{\,\bigtriangledown}$, 0.86$^{\,\bigtriangledown}$) 10$^{-3}$ &
 (6.5$^{\,\Diamond}$, 2.0$^{\,\Diamond}$) 10$^{-4}$ &
 (5.7$^{\,\Box}$, $\varnothing$) 10$^{-1}$ &
 (5.0$^{\,\Box}$, 1.5$^{\,\Box}$) 10$^{-4}$ & 
 (1.7$^{\,\Box}$, $\varnothing$) \\
\textbf{2333} \Tstrut\Bstrut & 
 (1.9$^{\,\bigtriangledown}$, 0.63$^{\,\bigtriangledown}$) 10$^{-1}$ &
 (2.0$^{\,\bigtriangledown}$, 0.65$^{\,\bigtriangledown}$) 10$^{-1}$ &
 (1.2$^{\,\bigtriangledown}$, 0.40$^{\,\bigtriangledown}$) 10$^{-1}$ &
 (1.2$^{\,\bigtriangledown}$, 0.42$^{\,\bigtriangledown}$) 10$^{-1}$ &
 (3.4$^{\,\bigtriangledown}$, 1.0$^{\,\bigtriangledown}$) 10$^{-2}$ &
 (3.1$^{\,\bigtriangledown}$, 1.0$^{\,\bigtriangledown}$) 10$^{-2}$ &
 (9.1$^{\,\bigtriangledown}$, 3.0$^{\,\bigtriangledown}$) 10$^{-2}$ &
 (9.4$^{\,\bigtriangledown}$, 3.1$^{\,\bigtriangledown}$) 10$^{-2}$ \\
\textbf{3311} \Tstrut\Bstrut & 
 $\varnothing$ &
 2.7$^{\,\Box}$ 10$^{-1}$ &
 $\varnothing$ &
 5.8$^{\,\Box}$ 10$^{-2}$ &
 $\varnothing$ &
 $\varnothing$ &
 $\varnothing$ &
 $\varnothing$ \\
\textbf{3312} \Tstrut\Bstrut & 
 (8.0$^{\,\Box}$, 3.4$^{\,\Box}$) 10$^{-4}$ &
 (4.5$^{\,\Box}$, 2.0$^{\,\Box}$) 10$^{-7}$ &
 (12$^{\,\Box}$, 5.0$^{\,\Box}$) 10$^{-5}$ &
 (9.6$^{\,\Box}$, 8.1$^{\,\Box}$) 10$^{-8}$ &
 (2.7$^{\,\Diamond}$, 6.1$^{\,\Diamond}$) 10$^{-5}$ &
 ($\varnothing$, $\varnothing$) & 
 (6.6$^{\,\Diamond}$, 15$^{\,\Diamond}$) 10$^{-6}$ &
 ($\varnothing$, $\varnothing$) \\
\textbf{3313} \Tstrut\Bstrut & 
 (1.0$^{\,\triangle}$, 1.0$^{\,\triangle}$) 10$^{-3}$ &
 (3.3$^{\,\triangle}$, 3.2$^{\,\triangle}$) 10$^{-5}$ &
 (6.4$^{\,\triangle}$, 4.8$^{\,\triangle}$) 10$^{-4}$ & 
 (1.9$^{\,\triangle}$, 1.4$^{\,\triangle}$) 10$^{-5}$ &
 (1.8$^{\,\Diamond}$, 4.0$^{\,\Diamond}$) &
 ($\varnothing$, $\varnothing$) & 
 ($\varnothing$, 11$^{\,\Diamond}$) & 
 ($\varnothing$, $\varnothing$) \\
\textbf{3322} \Tstrut\Bstrut & 
 $\varnothing$ &
 2.7$^{\,\Box}$ 10$^{-1}$ &
 $\varnothing$ &
 5.8$^{\,\Box}$ 10$^{-2}$ &
 $\varnothing$ &
 $\varnothing$ &
 $\varnothing$ &
 $\varnothing$ \\
 \textbf{3323} \Tstrut\Bstrut & 
 (5.3$^{\,\bigtriangledown}$, 1.8$^{\,\bigtriangledown}$) 10$^{-3}$ &
 (1.7$^{\,\bigtriangledown}$, 0.58$^{\,\bigtriangledown}$) 10$^{-4}$ &
 (3.6$^{\,\bigtriangledown}$, 1.2$^{\,\bigtriangledown}$) 10$^{-3}$ & 
 (1.1$^{\,\bigtriangledown}$, 0.37$^{\,\bigtriangledown}$) 10$^{-4}$ &
 (1.5$^{\,\Diamond}$, 1.8$^{\,\Diamond}$) 10$^{-2}$ &
 (8.7$^{\,\bigtriangledown}$, 2.9$^{\,\bigtriangledown}$) & 
 (1.1$^{\,\Diamond}$, 1.3$^{\,\Diamond}$) 10$^{-2}$ &
 ($\varnothing$, 8.7$^{\,\bigtriangledown}$) \\
\textbf{3333} \Tstrut\Bstrut & 
 $\varnothing$ &
 4.5$^{\,\bigtriangledown}$  &
 $\varnothing$ &
 2.9$^{\,\bigtriangledown}$  &
 $\varnothing$ &
 7.3$^{\,\bigtriangledown}$ 10$^{-1}$ &
 $\varnothing$ &
 2.2$^{\,\bigtriangledown}$ \\ [1.1ex]
\hline
\end{tabular}}
  \caption{Same as table \ref{tab:HQ} for $C^{Qd^{(1)}}_{ijkl}$, $C^{Qd^{(8)}}_{ijkl}$, $C^{Qu^{(1)}}_{ijkl}$ and $C^{Qu^{(8)}}_{ijkl}$.}
 \label{tab:QdQu}
 \vspace{-3mm}
\end{table*}

\begin{table*}[t!]
\resizebox{\textwidth}{!}{
\begin{tabular}{|c|c|c|c|c|c|c|c|c|}
\hline
{\boldmath$O^{HQ^{(1[3])}}_{jk}$} 
& {\boldmath$O^{LedQ}_{jjkl}$}
& {\boldmath$O^{LeQu}_{jjkl}$}
& {\boldmath$O^{ud^{(1[8])}_{jklm}}$}
 & {\boldmath$O^{QuQd^{(1[8])}}_{jklm}$}
\Tstrut\Bstrut  \\ [1.01ex]
$ \big( H^{\dagger} i \overset{ \ \leftrightarrow_{\,[A]}}{D_{\mu}}H \big) \left( \bar{Q}_{j} \gamma^{\mu} \left[\tau^{A}\right] Q_{k} \right)$ \Tstrut\Bstrut 
&  $ \big( \bar{L}_{j} e_{j} \big) \left(\bar{d}_{k} Q_{l} \right)$ \Tstrut\Bstrut  
&   $ \big( \bar{L}_{j} e_{j} \big) i \tau^{2} \left(\bar{Q}_{k} u_{l} \right)$ \Tstrut\Bstrut 
&  $ \big( \bar{u}_{j} \gamma_{\mu} [T^{a}] u_{k} \big) \left(\bar{d}_{l} \gamma^{\mu} [T^{a}] d_{m} \right)$ \Tstrut\Bstrut 
&  $ \big( \bar{Q}_{j} \gamma_{\mu} [T^{a}] u_{k} \big) i \tau^{2} \left(\bar{Q}_{l} \gamma^{\mu} [T^{a}] d_{m} \right)$ \Tstrut\Bstrut \\[1.1ex]
\hline
\hline
{\boldmath$O^{QQ^{(1[3])}}_{jklm}$} 
& {\boldmath$O^{uu}_{jklm}$} 
& {\boldmath$O^{dd}_{jklm}$} & 
{\boldmath$O^{Qd^{(1[8])}}_{jklm}$} 
& {\boldmath$O^{Qu^{(1[8])}}_{jklm}$}
\Tstrut\Bstrut  \\ [1.01ex]
$ \big( \bar{Q}_{j} \gamma_{\mu} [\tau^{A}] Q_{k} \big) \left(\bar{Q}_{l} \gamma^{\mu} [\tau^{A}] Q_{m} \right)$ \Tstrut\Bstrut 
&  $ \big( \bar{u}_{j} \gamma_{\mu}  u_{k} \big) \left(\bar{u}_{l} \gamma^{\mu}  u_{m} \right)$ \Tstrut\Bstrut  
&  $ \big( \bar{d}_{j} \gamma_{\mu}  d_{k} \big) \left(\bar{d}_{l} \gamma^{\mu}  d_{m} \right)$ \Tstrut\Bstrut 
&  $ \big( \bar{Q}_{j} \gamma_{\mu} [T^{a}] Q_{k} \big) \left(\bar{d}_{l} \gamma^{\mu} [T^{a}] d_{m} \right)$ \Tstrut\Bstrut  
&  $ \big( \bar{Q}_{j} \gamma_{\mu} [T^{a}] Q_{k} \big) \left(\bar{u}_{l} \gamma^{\mu} [T^{a}] u_{m} \right)$ \Tstrut\Bstrut  \\[1.1ex]
\hline
\end{tabular}}
\caption{Operators involved in the analysis. $H$, $Q$, and $L$ are Higgs, quark and lepton weak doublets; $d,u$ and $e$ quark and lepton $SU(2)_{L}$ singlets. $T^{a=1,\dots,8}$ are $SU(3)_{c}$ generators; $\tau^{A=1,2,3}$ $SU(2)_{L}$ ones. $\mu$ and $j,k,l,m$ are Lorentz and flavour indices.}
 \label{tab:SMEFT_Os}
 \vspace{-5mm}
\end{table*}

The most general $\Delta F = 2$ weak effective Hamiltonian (see
eqs.~(6)-(7) of \cite{Bona:2007vi} and also~\cite{Buchalla:1995vs,Ciuchini:1997bw}) matches at tree level
only five types of gauge-invariant operators \cite{Aebischer:2015fzz,Endo:2018gdn}:
\begin{flalign}
\label{eq:matching}
& C_{1}(\mu_{W})  = -
\left( C^{Q Q^{(1)}}(\mu_{W})+C^{Q Q^{(3)}}(\mu_{W}) \right)/\Lambda^{2} \nonumber \ , \\
& C_{1}'(\mu_{W})  =  - C^{q q }(\mu_{W})/\Lambda^{2}  \ , \\
& C_{4}(\mu_{W})  =  C^{Q q^{(8)}}(\mu_{W})/\Lambda^{2}  \nonumber \ , \\
& C_{5}(\mu_{W})   =  
\left( 2 \, C^{Q q^{(1)}}(\mu_{W})-\frac{1}{3} C^{Q q^{(8)}}(\mu_{W})
                     \right)/\Lambda^{2}  \nonumber  \ . &&
\end{flalign}
For our notation on the SMEFT, see table~\ref{tab:SMEFT_Os} and
also~\cite{Dedes:2017zog,Celis:2017hod}. $Q$ and $q = u, d$ represent
$SU(2)_{L}$ gauge doublet and singlets respectively, and
$\mu_{W} \simeq M_{W}$ is the matching scale. Flavour indices in
eq.~\eqref{eq:matching} have been understood.  The Wilson coefficients
of the SMEFT evolve according to coupled RG equations that we solve at
lowest order keeping the logarithmic term:
\begin{equation}
\label{eq:RGflow}
C_A(\mu_W) \simeq \left(\delta_{AB} - \beta_{AB} \mathcal{R}\right) C_B(\Lambda)\, ,
\end{equation}
where $A$ and $B$ are generalized indices running over the flavour
structure and field content of the effective theory, and $\beta_{AB}$
characterizes the quantum mixing of $B$ into
$A$. Eqs.~\eqref{eq:matching}-\eqref{eq:RGflow} allow us to translate
existing bounds of the form $(X^{(\textrm{R})}, X^{(\textrm{I})})$
on the $\Delta F = 2$ operators of the weak effective theory, see table~\ref{tab:DF2}, 
into constraints on the SMEFT, barring accidental cancellations:
\begin{flalign}
\label{eq:re_im_bounds}
\vert C_B^{(\textrm{R})}(\Lambda) \vert  <   \min \left( \frac{\Lambda^{2}X_{A}^{(\textrm{R})}}{\vert\delta_{AB} -  \beta_{AB}^{(\textrm{R})} \mathcal{R}\vert}, \frac{\Lambda^{2}X_{A}^{(\textrm{I})}}{\vert\ \beta_{AB}^{(\textrm{I})} \mathcal{R}\vert} \right)  ,    \\
\vert C_B^{(\textrm{I})}(\Lambda) \vert  <   \min \left( \frac{\Lambda^{2}X_{A}^{(\textrm{I})}}{\vert\delta_{AB} -  \beta_{AB}^{(\textrm{R})} \mathcal{R}\vert}, \frac{\Lambda^{2}X_{A}^{(\textrm{R})}}{\vert \beta_{AB}^{(\textrm{I})} \mathcal{R}\vert} \right)  ,  \nonumber
\end{flalign}
where $(\textrm{R,I})$ superscripts denote real and imaginary
parts. While the bounds obtained through eq.~(\ref{eq:re_im_bounds}),
\emph{i.e.} switching on the real or the imaginary part of a single
Wilson coefficient at a time, are generally valid, a model-dependent
analysis becomes mandatory whenever two or more Wilson coefficients
(including real and imaginary parts) are close to the bounds we
provide, since interference effects might become important in that
case.

A few remarks are in order. The state-of-the-art knowledge on the
anomalous dimensions in eq.~\eqref{eq:RGflow} encodes all leading
order effects in the Higgs self-coupling, Yukawa insertions and SM
gauge couplings,
see~\cite{Jenkins:2013zja,Jenkins:2013wua,Alonso:2013hga,Alonso:2014zka}.
For our purpose, it suffices to consider the RG effects induced by the
Yukawa matrices, including also their (non-negligible) change from the
high scale $\Lambda$ to $\mu_{W}$, and by the strong gauge
coupling. Entries for the CKM matrix are taken from the NP fit of
\cite{UTfit2018}, neglecting small corrections expected in the SMEFT
context \cite{Descotes-Genon:2018foz} which would produce effects of
$\mathcal{O}(\Lambda^{-4})$ in our analysis. All other SM parameters
are defined at the EW scale following \cite{Xing:2007fb,Xing:2011aa}.  In order to
accurately study CP violation effects as well, we developed a
numerical code enabling us to fully explore the case of complex-valued
Wilson coefficients. The starting point for our numerical analysis is
the update \cite{UTfit2018} of the study in~\cite{Bona:2007vi}, which
provides bounds on the coefficients of the weak effective Hamiltonian.
The $(X^{(\textrm{R})}, X^{(\textrm{I})})$ bounds we present are
obtained from the (symmetrized) 95$\%$ probability interval per single
Wilson coefficient of the $\Delta F = 2$ weak Hamiltonian at
$\mu_{W}$.  We run up to the NP scale $\Lambda=1$ TeV and present
bounds on the SMEFT Wilson coefficients at that scale.  We discard
values larger than $4 \pi$ for $\Lambda=1$ TeV. One can convert a
bound $C_{A}<X$ at 1 TeV into a bound on $\Lambda=1/\sqrt{X}$ for
$C_{A}=1$, up to terms of $\mathcal{O}(\log (\Lambda/\mathrm{TeV}))$.

In table~\ref{tab:HQ} we collect
the bounds on the coefficients of operators involving Higgs and
left-handed quark bilinears. Comparing with~\cite{deBlas:2016nqo,Almeida:2018cld}, our bounds on
non-flavour-universal coefficients are two to three orders of
magnitude stronger than the ones on flavour-universal operators coming
from EW precision data and Higgs physics.  Remarkably, what reported
in table~\ref{tab:HQ} is a pure outcome of the RG flow in the context
of the SMEFT. The same is true for table \ref{tab:LE}, where we
present the constraints on the SMEFT dimension-six operator involving
anti-symmetric contractions of quark and lepton weak doublets,
together with quark and charged lepton singlets. The operator
$O^{LeQu}$ mixes into left-right $\Delta F=2$ operators in the up
sector via the up-quark Yukawa matrix and the diagonal lepton one, $Y_{\ell}$. Consequently, it gets constrained only by
$D-\bar{D}$ mixing in the basis where $Y_{D}$ is diagonal and $Y_{U}$ can generate flavour mixing. 
Conversely, operator $O^{LedQ}$ mixes into
left-right $\Delta F=2$ operators in the down sector via $Y_{\ell}$ and
$Y_{D}$, getting constrained only in the basis of diagonal
$Y_{U}$. In table \ref{tab:ud} we present bounds on four-quark
operators involving both up- and down-type singlets, which also mix
into flavour-violating $\Delta F=2$ operators in the down and up
sectors only via the operator mixing generated by $Y_U$ and $Y_D$
respectively. Another class of operators contributing to
$\Delta F=2$ processes only through mixing via Yukawa interactions is
given by four-quark operators involving two doublets, an up-type
singlet and a down-type singlet. The set of bounds found is reported in table \ref{tab:QuQd}.

Let us turn now to the operators
appearing in eq.~\eqref{eq:matching}. Depending on the
flavour indices, we may handle two different situations: \textit{i)} the operator matches directly onto the $\Delta F=2$ operators at the EW
scale, after going to the mass eigenstate basis; \textit{ii)} the operator mixes with the $\Delta F=2$ ones only via RG flow. Note that for \textit{i)} one still needs to keep track of RG effects inducing a
misalignment of the Yukawa couplings between the cutoff and the EW
scale. We stress here how $D-\bar{D}$ and $K-\bar{K}$ mixings play a crucial role in constraining operators involving quark left-handed doublets. CKM misalignment between $Y_U$ and $Y_D$ ensures indeed a non-vanishing contribution
of operators $O^{QQ^{(1,3)}}$ either to $\Delta S=2$ or $\Delta C=2$ transitions, leading to stringent
constraints from at least one of the two processes. The strongest constraints on $O^{dd}$ come instead from purely right-handed operators and consequently do not depend on the alignment in flavour space once the aforementioned  effects from Yukawa running are taken into account. Bounds for operators $O^{QQ^{(1,3)},dd}$ are in table~\ref{tab:QQuudd} and are strictly related to \textit{i)}. For $O^{uu}$ constraints follow mainly from \textit{ii)} and we find: $C^{uu}_{1112}< (\varnothing,6.1^{\,\Diamond})$ TeV$^{-2}$ and $C^{uu}_{1213}< (2.1^{\,\Diamond},0.065^{\,\Diamond})$ TeV$^{-2}$ for diagonal $Y_D$, $C^{uu}_{1213}< (\varnothing,2.2^{\,\Diamond})$ TeV$^{-2}$ for diagonal $Y_{U}$; the only exception being $C^{uu}_{1212} < (28^{\,\Diamond},0.83^{\,\Diamond}) \, 10^{-8}$ TeV$^{-2}$ independent on the flavour alignment chosen as reasonably expected. Finally, for the sake of completeness, in table \ref{tab:QdQu} we collect the large set of constraints found for $O^{Qd^{(1,8)},Qu^{(1,8)}}$. The discussion of the $\Delta F = 2$ bounds on these operators  is analogous to the one already given for $O^{QQ^{(1,3)}}$.

The results derived so far represent a very serious challenge for
models with new sources of flavour violation, including those cases
where NP couples solely (or dominantly) to third generation
quarks. Our strong constraints can be considerably weakened in MFV
models. Within MFV scenarios, we constrain $O^{HQ^{(1,3)}}$,
$O^{QQ^{(1,3)}}, O^{Qu^{(1)}}$ and $O^{Qu^{(8)}}$ operators, probing
NP scales as low as 1.3, 9.9, 1.4 and 0.8 TeV respectively.  These may
be regarded as the weakest possible limits to date on heavy new
dynamics coupled to the quark sector.

\textbf{Acknowledgements}: L.S. is associated to the Dip. di Fisica,
Universit{\`a} di Roma ``La Sapienza''. M.V. is supported
by the NSF Grant No.~PHY-1620638. This project has received
funding from the European Research Council (ERC) under the European
Union's Horizon 2020 research and innovation program (grant agreement
n$^o$ 772369).

\bibliographystyle{apsrev4-1}
\bibliography{main.bib}

\end{document}